\documentclass[twocolumn,english,prlb,twocolumn,floatfix,footinbib,superscriptaddress,showpacs, showkeys,notitlepage]{revtex4-1}
\usepackage[T1]{fontenc}
\usepackage[utf8x]{inputenc}
\usepackage{amsmath}
\usepackage{amssymb}
\usepackage{graphicx}
\usepackage{esint}

\makeatletter
\usepackage{babel}
 \usepackage[colorlinks]{hyperref}

\usepackage{babel}

\makeatother

\usepackage{babel}
\begin{document}

\title{Relaxation and decoherence of qubits encoded in collective states
of engineered magnetic structures}

\author{Alexey M. Shakirov}
\email[a.shakirov@rqc.ru]{}

\affiliation{Russian Quantum Center, Novaya street 100A, 143025 Skolkovo, Moscow
Region, Russia}

\affiliation{Department of Physics, Lomonosov Moscow State University, Leninskie
Gory 1, 119992 Moscow, Russia}

\author{Alexey N. Rubtsov}

\affiliation{Russian Quantum Center, Novaya street 100A, 143025 Skolkovo, Moscow
Region, Russia}

\affiliation{Department of Physics, Lomonosov Moscow State University, Leninskie
Gory 1, 119992 Moscow, Russia}

\author{Alexander I. Lichtenstein}

\affiliation{Institute of Theoretical Physics, Hamburg University, Hamburg D-20355,
Germany}

\author{Pedro Ribeiro}

\affiliation{CeFEMA, Instituto Superior Técnico, Universidade de Lisboa, Av. Rovisco
Pais, 1049-001 Lisboa, Portugal}
\begin{abstract}
The quantum nature of a microscopic system can only be revealed when
it is sufficiently decoupled from surroundings. Interactions with
the environment induce relaxation and decoherence that turn the quantum
state into a classical mixture. Here, we study the timescales of these
processes for a qubit encoded in the collective state of a set of
magnetic atoms deposited on a metallic surface. For that, we provide
a generalization of the commonly used definitions of $T_{1}$ and
$T_{2}$ characterizing relaxation and decoherence rates. We calculate
these quantities for several atomic structures, including a collective
spin, a setup implementing a decoherence-free subspace, and two examples
of spin chains. Our work contributes to the comprehensive understanding
of the relaxation and decoherence processes and shows the advantages
of the implementation of a decoherence free subspace in these setups.
\end{abstract}


\maketitle
\global\long\def\ket#1{\left| #1\right\rangle }

\global\long\def\bra#1{\left\langle #1 \right|}

\global\long\def\kket#1{\left\Vert #1\right\rangle }

\global\long\def\bbra#1{\left\langle #1\right\Vert }

\global\long\def\braket#1#2{\left\langle #1\right. \left| #2 \right\rangle }

\global\long\def\bbrakket#1#2{\left\langle #1\right. \left\Vert #2\right\rangle }

\global\long\def\av#1{\left\langle #1 \right\rangle }

\global\long\def\tr{\text{tr}}

\global\long\def\Tr{\text{Tr}}

\global\long\def\pd{\partial}

\global\long\def\im{\text{Im}}

\global\long\def\re{\text{Re}}

\global\long\def\sgn{\text{sgn}}

\global\long\def\Det{\text{Det}}

\global\long\def\abs#1{\left|#1\right|}

\global\long\def\up{\uparrow}

\global\long\def\down{\downarrow}

\global\long\def\k{\mathbf{k}}

\global\long\def\wks{\mathbf{\omega k}\sigma}

\global\long\def\vc#1{\mathbf{#1}}

\global\long\def\bs#1{\boldsymbol{#1}}

\global\long\def\t#1{\text{#1}}

\section{Introduction \label{sec:Introduction}}

Exploiting quantum dynamics for information processing, data storage,
or sensing requires us to manipulate and measure quantum states before
relaxation and decoherence processes set in. The characteristic timescales
associated with the energy relaxation and the loss of quantum coherences
depend crucially on the interaction with the local environment. In
the case of single magnetic atoms or molecules in contact with a metallic
surface \cite{Heinrich2004,Hirjibehedin2007,Tsukahara2009,Gauyacq2012},
the most relevant interaction is the magnetic exchange with itinerant
electrons in the substrate \cite{Delgado2010a,Delgado2014}. Additionally,
the magnetic degrees of freedom couple to substrate phonons and nuclear
spins \cite{Delgado2017}.

The ability to manipulate and address individual atoms and to perform
spin and time resolved measurements \cite{Stipe1998,Meier2008,Wiesendanger2009,Loth2010,Baumann2015,Krause2016,Paul2016,Lado2016,Yan2017}
permitted us to engineer a number of prototype setups \cite{Hirjibehedin2006,Otte2009,Spinelli2014,Choi2016,Girovsky2017}
and demonstrated the potential of these artificial magnetic structures
for classical information processing \cite{Imre2006,Khajetoorians2011}.
However, the proximity with the metallic substrate induces the creation
of electron-hole pairs that rapidly decohere the magnetic state \cite{Gauyacq2015,Delgado2015}
thus posing considerable limitations to the exploitation of the quantum
regime \cite{Leuenberger2001,Troiani2005,Bogani2008}. For example,
in 1D chains of Fe atoms coupled antiferromagnetically, experiments
observe activated switching between the two doubly degenerate (classical)
Néel states rather then the (quantum) nonmagnetic ground state \cite{Loth2012}.

One strategy to protect the spin states is environment engineering.
This route was explored in Ref. \cite{Heinrich2013}, employing a
superconductor substrate that forbids quasiparticle creation for energies
below the superconducting gap. An alternative path is to engineer
artificial structures with intrinsically large decoherence times.
A systematic way to proceed is to design systems supporting so-called
decoherence-free subspaces \cite{Bacon2001,Lidar2003,Lidar2014} where
quantum information can be stored without loss. A recently developed
method, able to capture the coherent quantum dynamics of atomic magnets
\cite{Shakirov2016a} allows for the theoretical study of how to implement
this strategy. In contrast to previous descriptions in terms of rate
equations \cite{Fernandez-Rossier2009,Lorente2009,Delgado2010a,Ternes2015}
that only describe the dynamics of the populations, this approach
also models the quantum coherences and allows us to access the current
statistics measured by the STM tip.

In this paper we study relaxation and decoherence processes arising
in engineered atomic spin structures due to the contact with the metallic
surface and the STM tip. We show that the standard concepts of the
timescales $T_{1}$ and $T_{2}$ introduced for two-level systems,
such as a single atom with spin $1/2$, are not well-defined in the
generic case. In particular, they do not work when dynamics of coherences
and populations are coupled, e.g., due to the interaction with the polarized STM tip.
The standard definitions are also not extendable to multilevel
quantum systems, where the transitions from the states encoding a qubit to other eigenstates can contribute to the relaxation and decoherence.
To characterize the quantum dynamics in the generic
case, we propose new quantities that generalize the notions of $T_{1}$
and $T_{2}$ and that can extend to cases where a qubit is encoded
in a subspace of a larger Hilbert space. This new approach enables
us to determine the timescales for the degradation of stored quantum
information in different engineered structures such as spin chains
and collective spin models.
In general, there are three such timescales, but in most of the studied examples two of them coincide, in which case one can associate the two different quantities with relaxation and decoherence processes.
We study these timescales in several
atomic structures including the proposed decoherence-free subspaces
and investigate how these quantities depend on the temperature and
bias voltage applied to the STM tip. Our work shows that the implementation
of a decoherence free subspace based on four spin-$1/2$ atoms is
realistic. 

The paper is organized as follows. In Sec. \ref{sec:Model} we describe
the model and the method based on the effective master equation developed
in Ref. \cite{Shakirov2016a}. In Sec. \ref{subsec:Generalized-relaxation-and}
we identify the problem with the definition of $T_{1}$ and $T_{2}$
for a generic dissipative evolution and provide a suitable generalization.
These quantities are studied for different atomic structures in Sec.
\ref{subsec:Examples}. In Sec. \ref{sec:Conclusion} we discuss our
results, provide conclusions and state some of the open questions.
The Appendices are devoted to some technical aspects: Appendix \ref{sec:RateEquations}
shows how to obtain the rate equations for the populations once coherences
are disregarded, Appendix \ref{sec:Comparison} addresses the comparison
of the master equation used in this work with that of Ref. \cite{Shakirov2016a}
where additional simplifications were performed, and Appendix \ref{sec:BlochEquation}
shows how to obtain a Bloch-like equation for the evolution of the
magnetization for the case of a single spin-$1/2$ atom. 

\section{Model \& Method \label{sec:Model}}

\subsection{Model }

\begin{figure}[t]
\includegraphics[width=0.9\columnwidth]{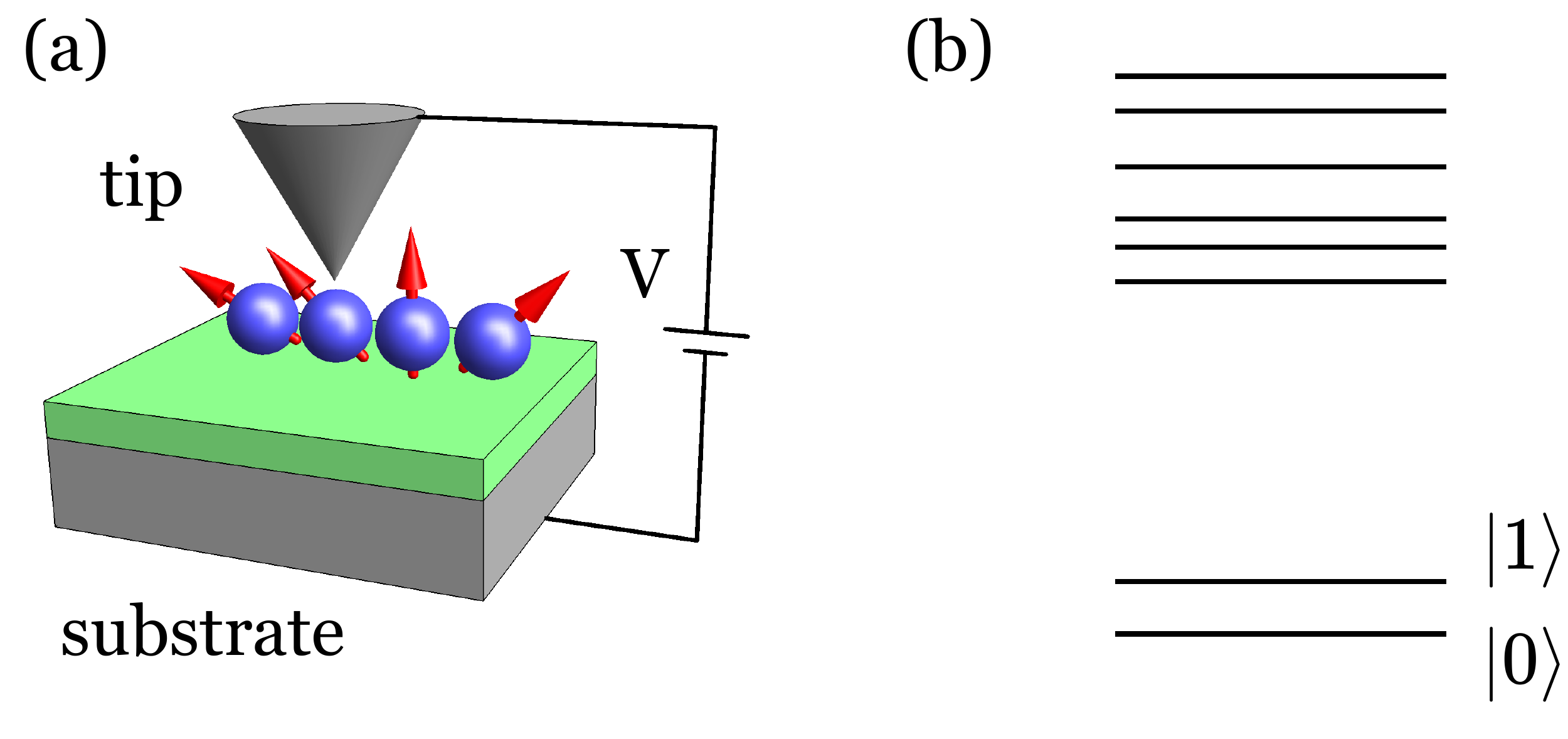}\caption{Schematic diagram of the model. (a) Sketch of an engineered atomic
spin device. Magnetic atoms are located on a metallic substrate coated
by an insulating layer, and a metallic tip is placed above one of
the atoms. (b) Energy spectra of an atomic structure used as a qubit.
Two lowest energy states are separated from the remaining part of
the spectra by an energy gap.}
\label{fgr:sketch}
\end{figure}

An engineered atomic spin device (EASD) consists of a cluster of magnetic
atoms arranged on a metallic substrate coated by a thin insulating
layer. Individual atomic degrees of freedom can be probed by placing
a spin polarized STM tip on top of the respective atom, as schematically
shown in Fig. \ref{fgr:sketch}. The entire system can be modeled
by a Hamiltonian $H=H_{S}+H_{E}+H_{I}$ that consists of three terms,
describing the atomic subsystem $H_{S}$, the electronic degrees of
freedom of the metallic leads (including the substrate and tip) $H_{E}$,
and the coupling between the atoms and these leads $H_{I}$.

The degrees of freedom of the atomic electrons reduce to those of
effective spins $S_{l}$, where $l=1\ldots L$ labels the atom, whenever
the Coulomb repulsion among electrons occupying localized orbitals
induces a large charge gap \cite{Anderson1966,Schrieffer1966}. In
the regime of weak hybridization between the localized atomic orbitals
and those of the itinerant electrons in the leads, tunneling of electrons
happens by virtual excitations of the atomic charge state. A suitable
low energy description can be obtained in terms of isolated magnetic
moments interacting with each other through an effective exchange
\cite{Gatteschi2006}. The interaction with the substrate further
induces an anisotropy in the $\hat{z}$ direction \cite{Hirjibehedin2007,Otte2008,Tsukahara2009}.
Finally, an external magnetic field $\boldsymbol{B}$ can be applied
to the system. The atomic Hamiltonian thus yields
\begin{align}
H_{S} & =\sum_{l}\left[DS_{lz'}^{2}+E\left(S_{lx'}^{2}-S_{ly'}^{2}\right)+\boldsymbol{h}\cdot\boldsymbol{S}_{l}\right]\nonumber \\
 & +\sum_{ll'}J_{ll'}\boldsymbol{S}_{l}\cdot\boldsymbol{S}_{l'},\label{eq:atoms}
\end{align}
where $S_{lx'}$, $S_{ly'}$, and $S_{lz'}$ are the atomic spin projections
along the hard, intermediate, and easy axis of the substrate. The Zeeman
terms with $\boldsymbol{h}=g\mu_{B}\boldsymbol{B}$ are proportional
to the atomic $g$ factor and to the Bohr magneton $\mu_{B}$.

We model the substrate as a set of $L$ identical leads that couple
independently to each atom. The tip is described as an extra lead
coupled to one of the atoms $\tilde{l}$. The Hamiltonian of the leads
is therefore given by
\begin{equation}
H_{E}=\sum_{\nu ks}\varepsilon_{\nu ks}c_{\nu ks}^{\dagger}c_{\nu ks}.\label{eq:leads}
\end{equation}
where $\nu=1,\ldots,L,t$ labels the leads, and electronic states
are characterized by momentum $k$ and spin $s$, quantized along
the tip polarization $\boldsymbol{P}$. The leads are in thermal equilibrium
with a common temperature $1/\beta$ (in energy units) and chemical
potentials $\mu_{s}=0$, $\mu_{t}=-eV$, where $V$ is the applied
voltage and $-e$ is the electron charge. They are further characterized
by the spin-polarized local densities of states $\varrho_{\nu s}\left(\varepsilon\right)=\mathcal{V}_{\nu}^{-1}\sum_{k}\delta\left(\varepsilon-\varepsilon_{\nu ks}\right)$,
where $\mathcal{V}_{\nu}$ is the corresponding volume. As the leads
are good metals, their description should be independent on specific
details, however, for definiteness, we use a density of states of
the form 
\begin{equation}
\varrho_{\nu s}(\varepsilon)=\frac{1}{2W}\left(1+p_{\nu}s\right)\Theta\left(|\varepsilon|-W\right),\label{eq:dos}
\end{equation}
where $W$ is the bandwidth, and $p_{\nu}$ is the polarization parameter
of the lead $\nu$ that may take values from $0$ to $1$. In the
following we assume that the substrate is nonmagnetic, i.e., $p_{s}=0$,
and the tip is polarized with $p_{t}=p$. We normalize the densities
of states by the condition $\sum_{s}\int\varrho_{\nu s}(\varepsilon)d\varepsilon=2$
that helps us to eliminate the problems with the logarithmic terms
in the master equation.

The coupling of the leads to the atoms is described by the Hamiltonian
\cite{Appelbaum1967,Kim2004,Fernandez-Rossier2009}
\begin{align}
H_{I} & =\sum_{la\nu\nu'}\sqrt{J_{l\nu}^{a}J_{l\nu'}^{a}}S_{la}\otimes s_{a}^{\nu\nu'},\label{eq:coupling}\\
s_{a}^{\nu\nu'} & =\frac{1}{\sqrt{\mathcal{V}_{\nu}\mathcal{V}_{\nu'}}}\sum_{kk'ss'}c_{\nu ks}^{\dagger}\frac{\sigma_{ss'}^{a}}{2}c_{\nu'k's'},\label{eq:spindensity}
\end{align}
where $a=0,x,y,z$, the axis $z$ is aligned with the tip polarization,
and $\sigma^{a}$ are the Pauli matrices (with $\sigma^{0}=I$). The
terms with $a=0$ in Eq. (\ref{eq:coupling}) correspond to the elastic
tunneling of electrons between the leads mediated by the atoms, while
the terms with $a\neq0$ describe the Kondo interaction of each atomic
spin with the spin density created by the lead electrons. The coupling
is assumed to be rotationally invariant, i.e., $J_{l\nu}^{a}=J_{l\nu}$
is the same for all $a\neq0$, and the coupling energies $J_{l\nu}\simeq2u_{l\nu}^{2}U_{l}\Delta_{l}^{-1}\left(\Delta_{l}+U_{l}\right)^{-1}$
are determined by the lead-atom hopping amplitudes $u_{l\nu}$, the
intra-atomic Coulomb repulsion energies $U_{l}$, and the gaps $\Delta_{l}$
between the atomic levels and the Fermi energy of the leads \cite{Schrieffer1966}.
Since each lead is coupled to only one atom, we use $J_{ll'}^{a}=J_{ls}^{a}\delta_{ll'}$
for the coupling to the substrate and $J_{lt}^{a}=J_{t}^{a}\delta_{l\tilde{l}}$
for the coupling to the tip. In the following we use dimensionless
parameters $\varGamma_{l\nu}^{a}=\pi J_{l\nu}^{a}\mathcal{V}_{\nu}/\left(2W\right)$
to characterize the strength of the lead-atom couplings.

\subsection{Master equation}

We investigate dynamics of an EASD by employing a master equation
\begin{equation}
\partial_{t}\rho=\mathcal{L}\rho\label{eq:master}
\end{equation}
for the density matrix of the atomic spins $\rho$, that is well adapted
to describe the regime of weak coupling between the atoms and the
leads. The standard approach to the construction of the superoperator
$\mathcal{L}$ invokes the Born and Markov approximations \cite{Breuer2002}.
If additionally the rotating wave approximation (RWA) is employed,
the superoperator $\mathcal{L}$ becomes of the Lindblad form. In
case the Hamiltonian of the isolated atom has a nondegenerate spectrum,
the latter approximation is equivalent to the neglect of the off-diagonal
elements of $\rho$, i.e., quantum coherences. Such a description based
on the rate equations for populations \cite{Delgado2010a,Ternes2015}
is suitable when decoherence in the system is much faster than relaxation
which is often the case in experiments due to hyperfine coupling \cite{LeGall2009}.
However, an account of coherences may become important in some situations \cite{Wichterich2007}.
In the context of EASDs, this happens for the case of nonparallel alignment of the tip polarization
and the magnetic fields acting on the atoms \cite{Shakirov2016a}.

In this paper, we do not employ the RWA and consider the coherent dynamics
of the setup given by the Redfield equation 
\begin{equation}
\pd_{t}\rho=-i\left[H'_{S},\rho\right]+\sum_{laa'}\left\{ \left[\varLambda_{laa'}\rho,S_{la}\right]+\textrm{\text{H.c.}}\right\} ,\label{eq:redfield}
\end{equation}
(the detailed derivation is given in Ref. \cite{Shakirov2016a}, with a minor difference discussed below), where the
Hamiltonian $H'_{S}=H_{S}+\Delta H_{S}$ is renormalized by terms
$\Delta H_{S}=pJ_{t}S_{\tilde{l}z}-\frac{\ln2}{2W}p^{2}J_{t}^{2}S_{\tilde{l}z}^{2}$
induced by the polarized tip. The dissipator is expressed through
operators $\varLambda_{laa'}$ given by
\begin{align}
\varLambda_{laa'} & =\sum_{\nu\nu'\alpha\alpha'}u_{laa'}^{\nu\nu'}\kappa\left(\omega_{\alpha}-\omega_{\alpha'}-\mu_{\nu}+\mu_{\nu'}\right)\nonumber \\
 & \times\ket{\alpha}\bra{\alpha}S_{la'}\ket{\alpha'}\bra{\alpha'},\label{eq:lambda}
\end{align}
where $\ket{\alpha}$ denotes the eigenstates of $H'_{S}$ with energies
$\omega_{\alpha}$, and 
\begin{align}
u_{laa'}^{\nu\nu'} & =\frac{1}{4\pi}\sqrt{\varGamma_{l\nu}^{a}\varGamma_{l\nu'}^{a}\varGamma_{l\nu}^{a'}\varGamma_{l\nu'}^{a'}}\nonumber \\
 & \times\tr\left[\left(1+p_{\nu}\sigma^{z}\right)\sigma^{a}\left(1+p_{\nu'}\sigma^{z}\right)\sigma^{a'}\right],\label{eq:ufactors-2}
\end{align}
\begin{align}
\kappa\left(\omega\right) & =\frac{g\left(\beta\omega\right)+i\,f\left(\beta\omega\right)}{\beta}-\frac{i}{\pi}\omega\ln\frac{\left|\omega\right|}{cW},\label{eq:kfunction}
\end{align}
where $g\left(x\right)=x/\left(e^{x}-1\right)$, $f\left(x\right)=\frac{1}{\pi}P\int dy\left[g\left(y\right)+y\,\Theta\left(-y\right)\right]/\left(x-y\right)$.
We note that the terms with $a=0$ or $a'=0$ in Eq. (\ref{eq:redfield})
vanish, i.e., the elastic tunneling does not affect dynamics of the
atoms.

The steady state $\rho_{\infty}$ of the system is determined by the
eigenstate of $\mathcal{L}$ with zero eigenvalue, i.e., $\mathcal{L}\rho_{\infty}=0$.
One may check that at zero voltage the steady state is given by the
Boltzmann distribution $\rho_{\infty}\propto e^{-\beta H_{S}}$ and
includes no coherences. At $V\neq0$ the excitations induced by the
current change the distribution, while the spin transfer torque \cite{Slonczewski1996,Delgado2010}
(in the case $p\neq0$) results in $\rho_{\infty}$ that is not diagonal
in the eigenbasis. The latter case cannot be captured by the standard
approach based on rate equations for the populations \cite{Delgado2010a,Ternes2015}.
These equations might be obtained from Eq. (\ref{eq:redfield}) by
substituting $\rho_{\alpha\alpha'}=p_{\alpha}\delta_{\alpha\alpha'}$
and are given in Appendix \ref{sec:RateEquations}.

In the presence of a voltage, an electric current is generated between
the tip and the substrate due to the coupling terms in Eq. (\ref{eq:coupling}).
Its average value can be expressed through the atomic density matrix
as
\begin{equation}
I=-e~\text{tr}\left(\mathcal{J}\rho\right)\label{eq:current}
\end{equation}
(see Ref. \cite{Shakirov2016a} for the detailed derivation) with the current
superoperator $\mathcal{J}$ defined as
\begin{align}
\mathcal{J}\rho & =\sum_{aa'}\left[J_{aa'}\rho S_{\tilde{l}a}+S_{\tilde{l}a}\rho J_{aa'}^{\dagger}\right],\label{eq:Jsuper}\\
J_{aa'} & =\sum_{\nu\nu'\alpha\alpha'}u_{\tilde{l}aa'}^{\nu\nu'}\kappa\left(\omega_{\alpha}-\omega_{\alpha'}-\mu_{\nu}+\mu_{\nu'}\right)\label{eq:joperators}\\
 & \times\left(\delta_{\nu t}-\delta_{\nu't}\right)\ket{\alpha}\bra{\alpha}S_{\tilde{l}a'}\ket{\alpha'}\bra{\alpha'}.\nonumber 
\end{align}
This current superoperator can be split into three parts \cite{Fernandez-Rossier2009,Delgado2010a}
\begin{align*}
\mathcal{J} & =\mathcal{J}_{e}+\mathcal{J}_{m}+\mathcal{J}_{i},
\end{align*}
where (i) $\mathcal{J}_{e}$ includes the terms of Eq. (\ref{eq:Jsuper})
with $a=a'=0$ and corresponds to the elastic component of the current,
(ii) $\mathcal{J}_{m}$ includes the terms of Eq. (\ref{eq:Jsuper})
with $a=0,\,a'\neq0$ and $a\neq0,\,a'=0$, and corresponds to the
magnetoresistive component of the current, and (iii) $\mathcal{J}_{i}$
includes the terms of Eq. (\ref{eq:Jsuper}) with $a\neq0,\,a'\neq0$,
and corresponds to the inelastic component of the current. The average
values of the first two components are given by
\begin{align}
I_{e} & =-e\,\tr\mathcal{J}_{e}\rho=g_{e}V,\\
I_{m} & =-e\,\tr\mathcal{J}_{m}\rho=2g_{m}p\langle S_{\tilde{l}z}\rangle V,
\end{align}
where $\langle S_{\tilde{l}z}\rangle=\tr\left(S_{\tilde{l}z}\rho\right)$,
and we defined the elastic conductance $g_{e}=\frac{1}{\pi}e^{2}\varGamma_{\tilde{l}s}^{0}\varGamma_{t}^{0}$,
the inelastic conductance $g_{i}=\frac{1}{\pi}e^{2}\varGamma_{\tilde{l}s}\varGamma_{t}$,
and the magnetoresistive conductance $g_{m}=\sqrt{g_{e}g_{i}}$.

We note that the master equation (\ref{eq:redfield}) contains terms
originating from the imaginary part of $\kappa\left(\omega\right)$,
see Eq. (\ref{eq:kfunction}), which were not considered in the previous
work \cite{Shakirov2016a}. The corresponding part of the superoperator
has eigenvalues with positive real part and may result in unphysical
dynamics when the density matrix loses its positivity. The physical
origin of this contribution is not fully understood, though it is
known that logarithmic terms in the bandwidth induce an additional
shift to the energy levels \cite{Breuer2002,Uchiyama2009,Oberg2014,Delgado2014}. It is tempting
to interpret these logarithmic terms in the master equation as a weak
coupling signature of Kondo physics. However, they vanish for the
$V=0$ steady state described by the Boltzmann distribution and thus
cannot be responsible for the Kondo effect. The results of the current
spectra calculated with two master equations, with and without the
imaginary part of $\kappa\left(\omega\right)$, are qualitatively
similar, as shown in Appendix \ref{sec:Comparison}.

\section{Results\label{sec:Results}}

In this section we study relaxation and decoherence in spin based
qubits whose dynamics may be described by the introduced model. We
pay special attention to the definitions of the corresponding timescales
$T_{1}$ and $T_{2}$ and revise them for the case when the qubit
is realized on quantum systems with more than two levels. We introduce
the generalized definitions for the relaxation and decoherence timescales
and use them to calculate $T_{1}$ and $T_{2}$ in some exemplary
systems. The discussion is lead in the context of EASDs but is easily
extendable to generic open quantum systems.

\subsection{Generalized relaxation and decoherence times \label{subsec:Generalized-relaxation-and}}

The notions of relaxation and decoherence times were first introduced
in the theory of nuclear magnetic resonance \cite{Bloch1946} and
have been applied to many open physical systems since then. In such
systems the interaction with surroundings causes the thermal equilibration
and loss of information. On a fundamental level, the first process
described by the relaxation time $T_{1}$ is explained by the redistribution
of the state populations due to the energy exchange with the surroundings.
The second process has a different timescale $T_{2}$ (typically much
shorter than $T_{1}$) and is due to the loss of quantum coherence.
These timescales turned out to be especially relevant in the context
of quantum information science where they are used to characterize
the quality of qubits. For two level systems, the times $T_{1}$ and
$T_{2}$ are uniquely defined. However, qubits are often realized
on quantum systems which have more than two states. Though the quantum
information in this case is still recorded in the subspace of two
relevant states, the transitions to other states induced by the surroundings
cannot be disregarded. Thus the notions of relaxation and decoherence
times for such qubits have to be clarified.

In this subsection we first show how the timescales $T_{1}$ and $T_{2}$
arise for a single atom with spin $S=1/2$ (two-level system) and
study their behavior as different parameters of the setup, such as
the voltage and the tip polarization, are tuned. We then discuss possible
generalized definitions for $T_{1}$ and $T_{2}$ that (i) work for
systems with more than two levels, (ii) are physically justified,
and (iii) recover the standard $T_{1}$ and $T_{2}$ for two-level
systems.

\subsubsection{Statement of the problem for spin $S=1/2$}

A single atom with spin $S=1/2$ is a two level system whose density
matrix has a standard geometrical representation as a Bloch vector.
The connection between the Bloch vector $\boldsymbol{p}$ (with $\left|\boldsymbol{p}\right|\leqslant1$
for physically meaningful states) and the density matrix $\rho$ (in
the eigenbasis of $H_{S}=-\frac{1}{2}h\sigma^{z}$) is given by the
relations
\begin{align}
\rho & =\frac{1}{2}\left(I+\boldsymbol{\sigma}\cdot\boldsymbol{p}\right),\label{eq:ptorho}\\
\boldsymbol{p} & =\tr\left(\boldsymbol{\sigma}\rho\right),\label{eq:rhotop}
\end{align}
where $I$ is the identity matrix and $\boldsymbol{\sigma}=\left\{ \sigma^{x},\sigma^{y},\sigma^{z}\right\} $
is the Pauli vector. The Redfield equation (\ref{eq:redfield}) for
$\rho$ maps to the Bloch equation of the form 
\begin{equation}
\partial_{t}\boldsymbol{p}=-\boldsymbol{\omega}\times\left(\boldsymbol{p}-\boldsymbol{p}_{0}\right)-\bs R.\left(\boldsymbol{p}-\boldsymbol{p}_{0}\right),\label{eq:Bloch}
\end{equation}
with explicit expressions for a vector $\boldsymbol{\omega}$, a real
symmetric matrix $\bs R$ and the steady state $\boldsymbol{p}_{0}$
given in Appendix \ref{sec:BlochEquation}.

We first consider the situation when the tip is not polarized and
no voltage is applied across the system, i.e., $p=0$ and $V=0$. In
this case the atom relaxes to the Boltzmann state with Bloch vector
$\boldsymbol{p}_{0}=\tanh\left(\beta h/2\right)\hat{z}$, and
\begin{align}
\boldsymbol{\omega} & =\left\{ h-\frac{1}{T_{\varphi}}\left[f\left(\beta h\right)-\frac{\beta h}{\pi}\ln\frac{h}{cW}\right]\right\} \hat{z},\label{eq:omegaatV0}
\end{align}
where we introduced $T_{\varphi}=2\pi\beta/\left(\varGamma_{s}+\varGamma_{t}\right)^{2}$.
The matrix $\boldsymbol{R}$ has the structure
\begin{equation}
\boldsymbol{R}.\boldsymbol{p}=\frac{1}{T_{1}}\boldsymbol{p}_{\parallel}+\frac{1}{T_{2}}\boldsymbol{p}_{\perp},\label{eq:Rmatrixdecoupled}
\end{equation}
so that relaxation of the Bloch vector components $\boldsymbol{p}_{\parallel}$
and $\boldsymbol{p}_{\perp}$ perpendicular and parallel to $\hat{z}$
is decoupled. The timescales $T_{1}$ and $T_{2}$ characterize the
processes of the population imbalance relaxation and the decoherence
correspondingly and are given by
\begin{align}
\frac{1}{T_{1}} & =\frac{1}{T_{\varphi}}\left[g\left(\beta h\right)+g\left(-\beta h\right)\right],\label{eq:T1atV0}\\
\frac{1}{T_{2}} & =\frac{1}{T_{\varphi}}\left[\frac{g\left(\beta h\right)+g\left(-\beta h\right)}{2}+1\right],\label{eq:T2atV0}
\end{align}
where $g\left(x\right)$ is defined below Eq. (\ref{eq:kfunction}).
One can see that $1/T_{2}=1/(2T_{1})+1/T_{\varphi}$, so that $T_{\varphi}$
has physical meaning of the pure dephasing time \cite{Cohen-Tannoudji1992,Delgado2017}.
Both timescales depend on the energy difference between two states
$\omega_{1}-\omega_{0}=h$, reaching maximum for $h=0$. They might
be associated with the eigenvalues of superoperator $\mathcal{L}$
from the master equation (\ref{eq:master}). For spin $S=1/2$ this
superoperator has one zero eigenvalue, $\lambda_{0}=0$, corresponding
to the steady state, and three nonzero eigenvalues $\lambda_{i=1,2,3}$,
with $\lambda_{1}$ being real and two others forming a pair of conjugates
$\lambda_{2}=\lambda_{3}^{*}$. They relate to the relaxation and
decoherence times as
\begin{align}
\lambda_{1} & =-1/T_{1},\label{eq:T1definition}\\
\lambda_{2,3} & =-1/T_{2}\pm i\omega.\label{eq:T2definition}
\end{align}

If unpolarized current is driven through the atom, i.e., $p=0$ and
$V\neq0$, the steady state $\boldsymbol{p}_{0}$ and $\boldsymbol{\omega}$
remain aligned with $\hat{z}$ but depart from their equilibrium magnitudes.
The relaxation matrix still has the structure (\ref{eq:Rmatrixdecoupled}),
but the timescales $T_{1}$ and $T_{2}$ differ from the equilibrium
ones (\ref{eq:T1atV0}) and (\ref{eq:T2atV0}) by
\begin{align*}
\frac{1}{T_{1}} & =\left.\frac{1}{T_{1}}\right|_{V=0}+\frac{\varGamma_{s}\varGamma_{t}}{2\pi\beta}\left[\tilde{g}\left(h,V\right)+\tilde{g}\left(-h,V\right)\right],\\
\frac{1}{T_{2}} & =\left.\frac{1}{T_{2}}\right|_{V=0}+\frac{\varGamma_{s}\varGamma_{t}}{2\pi\beta}\left[\frac{\tilde{g}\left(h,V\right)+\tilde{g}\left(-h,V\right)}{2}+\tilde{g}\left(0,V\right)\right],
\end{align*}
where we used $\tilde{g}\left(h,V\right)=g\left(\beta\left(h+eV\right)\right)+g\left(\beta\left(h-eV\right)\right)-2g\left(\beta h\right)>0$.
In this form it is evident that $T_{1}$ and $T_{2}$ decrease compared
to their zero bias values.

If polarized current is driven through the atom, i.e., $p\neq0$ and
$V\neq0$, both $\boldsymbol{\omega}$ and $\boldsymbol{p}_{0}$ are
affected by the spin transfer torque and, as soon as $\boldsymbol{P}\nparallel\boldsymbol{B}$,
are no longer aligned with $\hat{z}$ nor with each other. The dynamics
of the components $\boldsymbol{p}_{\parallel}$ and $\boldsymbol{p}_{\perp}$
of the Bloch vector get coupled, i.e., the relaxation matrix does not
have the form (\ref{eq:Rmatrixdecoupled}). In this case the standard
concepts of timescales $T_{1}$ and $T_{2}$ become ill defined and
have to be reconsidered. The possible solution is to define them through
Eqs. (\ref{eq:T1definition}) and (\ref{eq:T2definition}).
Thus introduced $T_{1}$ and $T_{2}$ are univocal for a two-level system whose superoperator $\mathcal{L}$ has one real nonzero eigenvalue and two eigenvalues forming a complex-conjugated pair.
In this case the decay of the density matrix corresponding to the eigenvalue (\ref{eq:T1definition}) mostly affects the populations, while the eigenvalues (\ref{eq:T2definition}) correspond to the decay process mainly involving coherences.
However, it might also happen that $\mathcal{L}$ has three distinct real eigenvalues \cite{Kimura2002,Sudarshan2003}, in which case such clear demarcation between relaxation and decoherence is not possible.

\subsubsection{Definition of generalized relaxation and decoherence times}

Our ability to introduce relaxation and decoherence times in the previous
subsection relied on the fact that the system had two levels, and
one was able to uniquely express $T_{1}$ and $T_{2}$ through the
eigenvalues of the superoperator $\mathcal{L}$. This is not possible
for spin systems with more than two levels. If two distinct eigenstates
of the system $\ket 0$ and $\ket 1$ are used to realize a qubit,
the timescales $T_{1}$ and $T_{2}$, by their physical meaning, should
be defined as decay times for $\rho_{00}-\rho_{11}$ and $\rho_{01}$
correspondingly. However, dynamics of these quantities is governed
by multiple timescales determined by the eigenvalues of $\mathcal{L}$.
This is due to the fact that transitions to other eigenstates may
happen, so that the state of the system does not remain in the subspace
spanned by $\ket 0$ and $\ket 1$, as was the case for two-level
systems.

We now introduce generalized definitions for the relaxation and decoherence
times given they must characterize the ability of a qubit to store
information about the initial state. For a system with the Hilbert
space of dimension $d$, the initial states are prepared within the
subspace $\left\{ \ket 0,\ket 1\right\} $ and can be written as $\rho_{\boldsymbol{v}}=\frac{1}{2}\left(I+\boldsymbol{v}\cdot\boldsymbol{\sigma}\right)$,
where $I$ and $\sigma^{x,y,z}$ are $d\times d$ matrices obtained
from corresponding $2\times2$ matrices by filling missing elements
with zeros. The distinguishability of two quantum states can be characterized
by the Hilbert-Schmidt distance between their density matrices $||\rho-\rho'||\equiv\sqrt{\tr\left[\left(\rho-\rho'\right)^{2}\right]}$.
This distance does not change over time in isolated systems but decreases
in open quantum systems whose density matrices evolve towards a steady
state. Therefore we consider the quantity
\begin{align}
D\left(\boldsymbol{\Delta},t\right) & \equiv\frac{||\rho_{\boldsymbol{v}}(t)-\rho_{\boldsymbol{v}'}(t)||}{||\rho_{\boldsymbol{v}}-\rho_{\boldsymbol{v}'}||},\label{eq:normdistance}
\end{align}
where $\boldsymbol{\Delta}=\boldsymbol{v}-\boldsymbol{v}'$ and $\rho_{\boldsymbol{v}}(t)=e^{\mathcal{L}t}\rho_{\boldsymbol{v}}$
is the time evolved density matrix. Our goal is to associate the generalized
$T_{1}$ and $T_{2}$ with the decay timescales of this quantity.
We first note that
\begin{align}
\rho_{\boldsymbol{v}}\left(t\right)-\rho_{\boldsymbol{v}'}\left(t\right) & =\frac{1}{2}\boldsymbol{\Delta}\cdot e^{\mathcal{L}t}\boldsymbol{\sigma}
\end{align}
depends only on the difference between vectors $\boldsymbol{v}$ and
$\boldsymbol{v}'$ which was used in Eq. (\ref{eq:normdistance}).
Treating matrices as vector states for the superoperator action and
using the notation $\rho\rightarrow\kket{\rho}$, one may decompose
the master equation superoperator as $\mathcal{L}=\sum_{i}\lambda_{i}\kket{\rho_{i}}\bbra{\tilde{\rho}_{i}}$,
where $\kket{\rho_{i}}$ and $\bbra{\tilde{\rho}_{i}}$ are eigenvectors
of $\mathcal{L}$ with eigenvalue $\lambda_{i}$, obtaining 
\begin{align}
\kket{\rho_{\boldsymbol{v}}\left(t\right)}-\kket{\rho_{\boldsymbol{v}'}\left(t\right)} & =\frac{1}{2}\sum_{ia}\bbrakket{\tilde{\rho}_{i}}{\sigma^{a}}\Delta_{a}e^{\lambda_{i}t}\kket{\rho_{i}},
\end{align}
where $a=x,y,z$. The distance between time evolved density matrices
then reads
\begin{align}
||\rho_{\boldsymbol{v}}(t)-\rho_{\boldsymbol{v}'}(t)|| & =\sqrt{\frac{1}{2}\sum_{aa'}M_{aa'}\left(t\right)\Delta_{a}\Delta_{a'}},
\end{align}
where we introduced a time-dependent $3\times3$ matrix $\boldsymbol{M}\left(t\right)$
with elements
\begin{align}
M_{aa'}\left(t\right) & =\frac{1}{2}\sum_{ii'}\bbrakket{\tilde{\rho_{i}}}{\sigma^{a}}\bbrakket{\tilde{\rho_{i'}}}{\sigma^{a'}}\label{eq:Mmatrix}\\
 & \times\tr(\rho_{i}\rho_{i'})e^{(\lambda_{i}+\lambda_{i'})t}.\nonumber 
\end{align}
Equation (\ref{eq:normdistance}) then transforms to

\begin{align}
D\left(\boldsymbol{\Delta},t\right) & =\sqrt{\frac{\boldsymbol{\Delta}^{T}.\boldsymbol{M}\left(t\right).\boldsymbol{\Delta}}{\boldsymbol{\Delta}^{T}.\boldsymbol{\Delta}}}.
\end{align}
From this expression one may see that the decay of $D\left(\boldsymbol{\Delta},t\right)$
has three timescales determined by the decay times of three eigenvalues
of $M\left(t\right)$. We denote these eigenvalues as $m_{k}\left(t\right)$,
$k=1,2,3$ and define timescales $T_{k}$ through the relations $\sqrt{m_{k}\left(T_{k}\right)}=1/e$,
using a (quasi)monotonic decrease of $m_{k}\left(t\right)$ from $1$
at $t=0$ to $0$ at $t=\infty$.

For most of the systems considered below, we found that two of these timescales
form a pair of close values $T_{2}\simeq T_{3}$ (differing by less than 1$\%$) and the third one
$T_{1}$ is different from them. In such cases we associate $T_{1}$ with relaxation and $T_{2}=T_{3}$ with decoherence, in analogy with two-level systems.
As shown in Appendix \ref{sec:TimesForTLS}, the introduced relaxation and decoherence times coincide with the standard definitions for two-level systems when dynamics of populations and coherences
are decoupled.
In case all three timescales $T_{1}$, $T_{2}$, and $T_{3}$ are distinct from each other, a clear identification of relaxation and decoherence processes is not possible as these timescales generically correspond to rates of information loss that affect both diagonal and off-diagonal elements of the density matrix.

\subsection{Relaxation and decoherence for different setups\label{subsec:Examples}}

In order to store and manipulate a qubit encoded in the quantum state
of an assembly of interacting magnetic atoms, the system is required
to have two low-lying energy states and relaxation and decoherence
times much larger than storage and manipulation timescales of the
qubit. In this subsection we use the introduced definitions to study
the properties of $T_{1}$ and $T_{2}$ in several systems engineered
to have two lowest energy states separated from the remaining part
of the spectrum, as shown in Fig. \ref{fgr:sketch}. We consider (i)
atomic systems that can be described by one spin variable and (ii)
atomic chains coupled by spin-spin interactions. We study the dependence
of the relaxation and decoherence times on the system size (spin $S$
for single atoms and length $L$ for spin chains) and on the parameters
of the environment such as the voltage, the temperature, and the external
magnetic field. We also discuss the strategies for decoherence suppression
in such devices. For Ising chains with longitudinal coupling, we show
three timescales $T_{1}$, $T_{2}$, and $T_{3}$ instead of two because
the identification of a pair of close values is not possible in this
case.

\subsubsection{Collective spin models}

\begin{figure}[t]
\includegraphics[width=1\columnwidth]{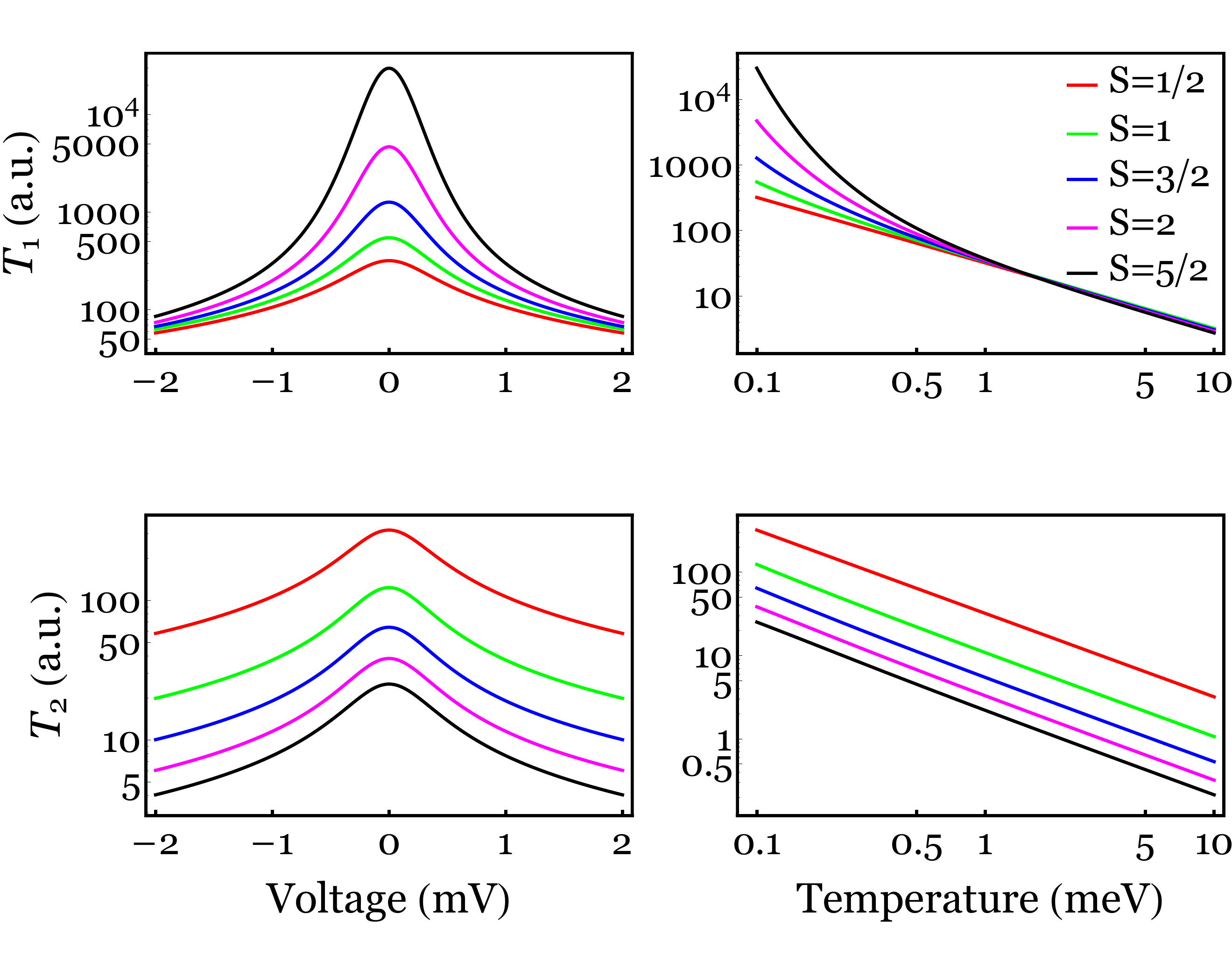}

\caption{Generalized relaxation and decoherence times $T_{1}$ and $T_{2}$
for single atoms of different spin $S$ as functions of the voltage
and the temperature. The tip is unpolarized, and we use the parameters
$D=-0.1$ meV, $J_{s}=J_{t}=1$ meV, and $W=10$ meV. For the voltage
curves the temperature is fixed at $k_{B}T=0.1$ meV, for the temperature
curves the voltage is fixed at zero.}
\label{fgr:SingleAtom}
\end{figure}

In many cases the Hamiltonian of the atomic structure commutes with
the square of the total spin $\boldsymbol{S}=\sum_{l}\boldsymbol{S}_{l}$,
and the ground state is either in the highest or lowest spin sector.
Typically, the highest multiplicity is more energetically favorable
due to the Hund's rule. Even if $S^{2}$ does not commute with the
dissipative terms in the Liouvilian, a description in terms of a collective
spin-$S$ is possible, assuming that the other spin sectors are much
higher in energy and cannot be excited by the environment. 

The situation described above amounts to consider a single spin-$S$
in Eq. (\ref{eq:atoms}). In order to have two low-lying states that
can encode a qubit, we consider $D<0$ such as to obtain an inverted
parabolic spectrum with a ground-state manifold spanned by $\ket 0=\ket{-S}$
and $\ket 1=\ket{+S}$ \cite{Khajetoorians2013,Miyamachi2013}. For
simplicity we consider the case $E=h=0$. The times $T_{1}$ and $T_{2}$
for the qubits realized on these states are presented in Fig. \ref{fgr:SingleAtom}
as functions of the voltage and the temperature for different values
of $S$. In the numerical calculations we assume an unpolarized tip
and use the parameters $D=-0.1$ meV for the anisotropy, $J_{s}=J_{t}=1$ meV for the inelastic coupling to the leads, and $W=10$
meV for the bandwidth of the leads. One observes that the relaxation time $T_{1}$ increases with
$S$, while the decoherence time $T_{2}$ shows the opposite behavior.
This can be explained by the fact that the lowest order transitions
induced by the environment can only change the atomic spin projection
by 1. Switching between the two ground states thus requires a series
of $2S$ transitions with the largest transition energy being $D\left(2S-1\right)$.
Two effects then lead to an increase of $T_{1}$ as $S$ increases:
More transitions are needed in the cascade, and the transition energy
gets larger. At low temperatures, the Boltzmann weights account for
the exponential growth of $T_{1}$, while at large $T$, the relaxation
time is inversely proportional to the temperature. The decoherence
time shows $1/T$ behavior in the whole temperature region. The dependence
of $T_{2}$ on the atomic spin is explained by the fact that the two
ground states have distinct magnetizations, and the larger they are,
the more effectively spin fluctuations destroy the coherence between
them. The effect of the voltage in the reduction of both $T_{1}$
and $T_{2}$ is explained by the fact that higher energy electrons,
inducing faster transitions and larger spin fluctuations, become accessible
by inelastic scattering. In the large voltage limit, both the relaxation
and decoherence times decrease as $1/V$ which is simply explained
by the asymptotic behavior of the function $\kappa\left(\omega\right)$
in Eq. (\ref{eq:lambda}).

\begin{figure}

\includegraphics[width=1\columnwidth]{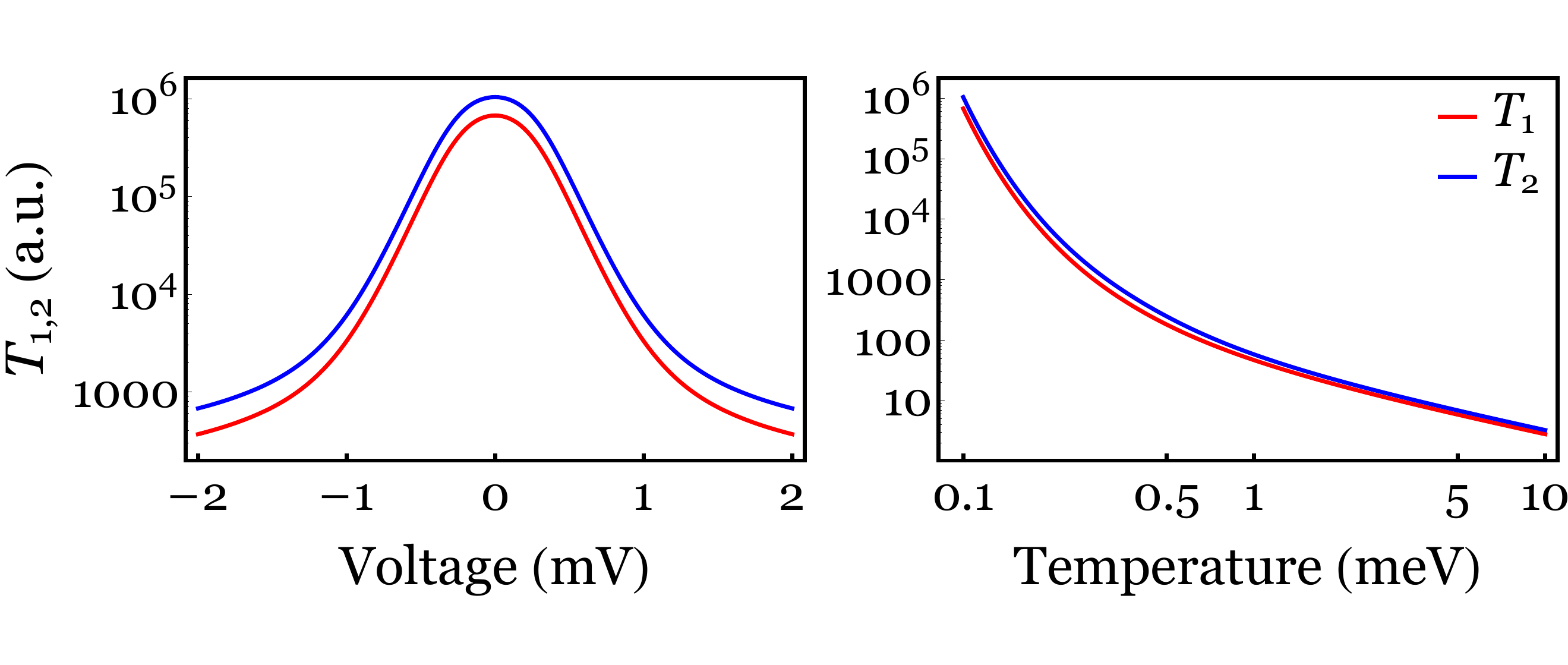}\caption{Generalized relaxation and decoherence times $T_{1}$ and $T_{2}$
for the collective spin of four pairwise coupled $S=1/2$ atoms as
functions of the voltage and the temperature. The tip is unpolarized,
and we use the parameters $J=0.5$ meV, $J_{s}=J_{t}=1$ meV, and
$W=10$ meV. For the voltage curves the temperature is fixed at $k_{B}T=0.1$
meV, for the temperature curves the voltage is fixed at zero.}
\label{fgr:CollectiveSpin}
\end{figure}

Let us now consider a different situation where the Hund's rule is
not satisfied, and the lowest energy states are singlets. The first
nontrivial case arises for a system of four $S=1/2$ atoms where the
two singlets are formed. Since the matrix elements of these states
with the environment coupling terms vanish, the evolution within the
singlet subspace is purely unitary. Such decoherence-free subspace was exploited in Ref. \cite{Bacon2001}
to realize a qubit robust to local decoherence.
However, while the leading order decoherence effects of the considered system are suppressed, finite values of $T_{1}$ and $T_{2}$ are still expected due to the presence of the other spin sectors.

Such a system can be realized setting $J_{ll'}=J>0$ in Eq. (\ref{eq:atoms})
with four $1/2$-spins, in which case the ground state is doubly degenerate
and comprised of two singlets $\ket 0$ and $\ket 1$. One may realize
such a Hamiltonian with four atoms coupled to each other pairwise with
the same Heisenberg energy, i.e., in the tetrahedron geometry. The
times $T_{1}$ and $T_{2}$ for the qubit encoded within the singlet
states are shown in Fig. \ref{fgr:CollectiveSpin} as functions of
the voltage and the temperature. In the calculations we assume the
unpolarized tip and use the parameters $J=0.5$ meV for the Heisenberg energy, $J_{s}=J_{t}=1$
meV for the inelastic coupling to the leads, and $W=10$ meV for the bandwidth of the leads. Note that, in this system, the relaxation and
decoherence times are of the same order in the whole region of parameters.
This means that pure dephasing is absent, and the finite value of
$T_{2}$ is only due to the transitions between the singlet states
mediated by the higher spin sectors whose states have higher energy.
Thus in the limit $T\to0$, due to the Boltzmann factor, both relaxation
and decoherence are exponentially suppressed. This should be contrasted
with the case of a single spin where $T_{2}\propto1/T$. The absence
of pure dephasing can be explained by the fact that all matrix elements
of local spin operators in the ground state subspace are zero, i.e.,
$\bra iS_{la}\ket j=0$ for $i,j=\ket 0,\ket 1$, and the right hand
side of the master equation projected onto this subspace vanishes
identically.

\subsubsection{Spin chains}

We now study the properties of the relaxation and decoherence times
for two types of $S=1/2$ spin chains: (i) Ising chains and (ii)
transverse field Ising chains with longitudinal coupling. For both
of them we assume that the tip is coupled to the atom at the end of
the chain, i.e., $\tilde{l}=1$.

\begin{figure}[t]

\includegraphics[width=1\columnwidth]{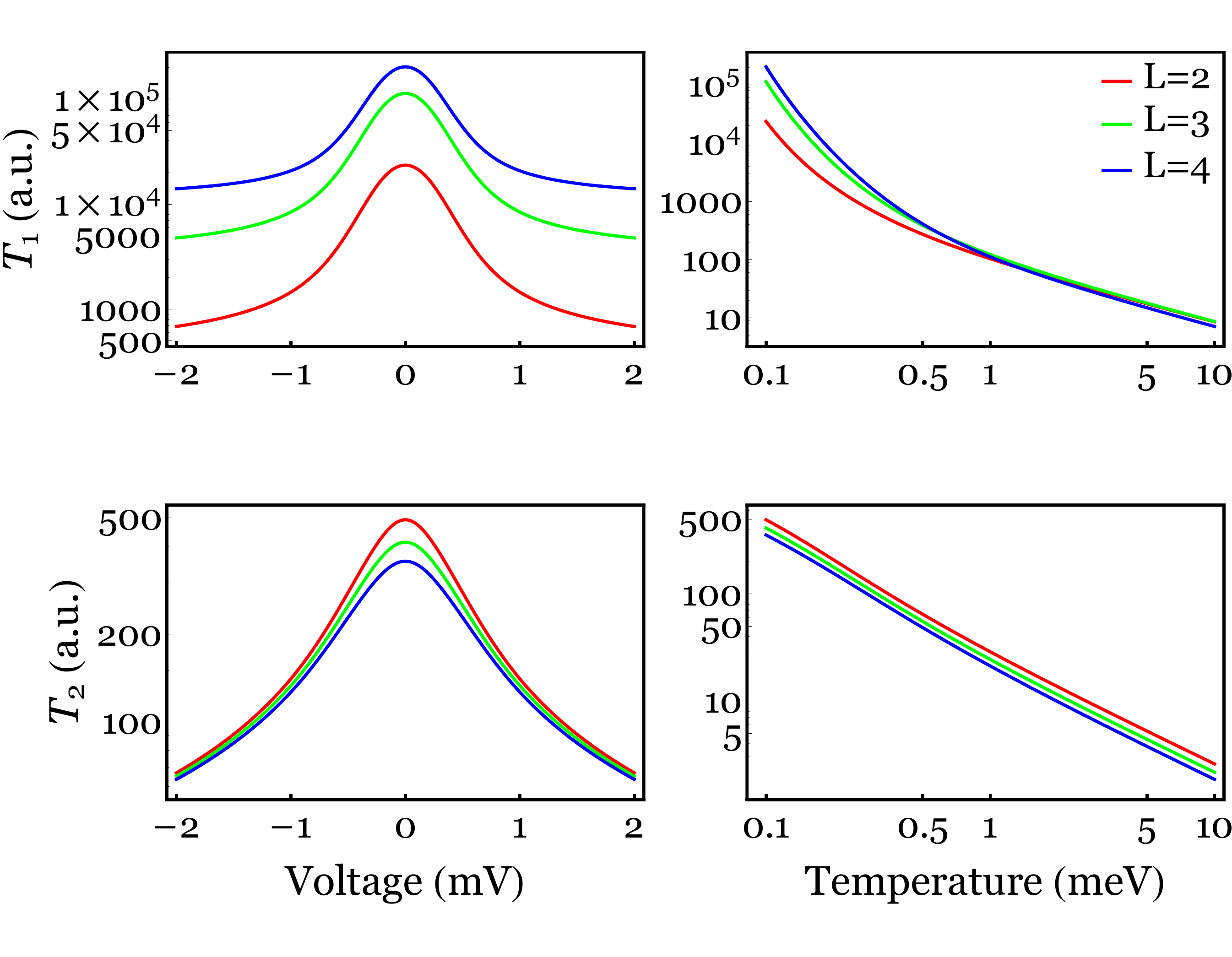}\caption{Generalized relaxation and decoherence times $T_{1}$ and $T_{2}$
for ferromagnetic Ising chains of different length $L$ as functions
of the voltage and the temperature. The tip is unpolarized, and we
use the parameters $J=-1$ meV, $J_{s}=J_{t}=1$ meV, and $W=10$
meV. For the voltage curves the temperature is fixed at $k_{B}T=0.1$
meV, for the temperature curves the voltage is fixed at zero.}
\label{fgr:FMIsingChain}
\end{figure}

The Ising chain is described by the Hamiltonian
\begin{align}
H_{S} & =J\sum_{l=1}^{L-1}S_{lz}S_{l+1,z},
\end{align}
where atoms have spin $1/2$ and $L$ is the chain length. The system
has doubly degenerate ground states which are (anti)ferromagnetically
ordered for $J<0$ ($J>0$). The relaxation and decoherence times
for the qubit realized on these states are presented in Fig. \ref{fgr:FMIsingChain}
as functions of the voltage and the temperature for the chains of
different length. In the calculations we assume the unpolarized tip
and use the parameters $J=-1$ meV, $J_{s}=J_{t}=1$ meV for the inelastic coupling to the leads, and $W=10$
meV for the bandwidth of the leads. One can observe that $T_{1}$ increases with the chain length,
while $T_{2}$ shows the opposite behavior. However, the dependence
of the decoherence time on the system size is not as pronounced as
in the case of single atoms of different total spin $S$ considered
above. This is because the energy barriers for the excitation processes
do not scale with $L$, and the transition between the two ground
states is realized by the excitation of a spin at the chain edge and
the diffusive propagation of the domain wall to the opposite edge.
Applying a voltage or increasing the temperature results in the decrease
of $T_{1}$ and $T_{2}$. Both timescales behave as $1/T$ and $1/V$
in the limits of large temperature and large voltage correspondingly.
Similarly to the case of single atoms, the relaxation time shows the
exponential growth at low temperature, while the decoherence time
preserves $1/T$ behavior in the whole temperature region. The results
are in qualitative accordance with experimental data from Ref. \cite{Loth2012}
where the properties of the switching rate (we associate it with $1/T_{1}$)
between two lowest energy states in spin chains were studied.

\begin{figure}[t]
\includegraphics[width=1\columnwidth]{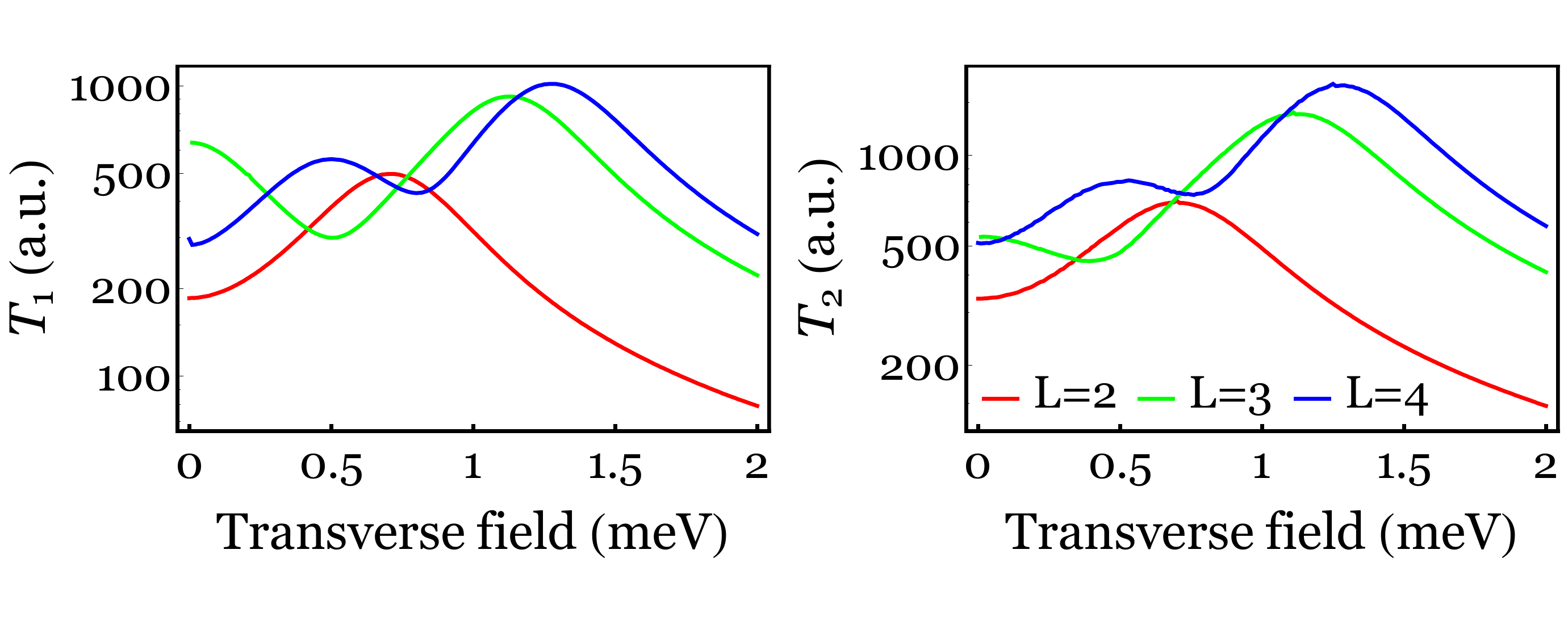}\caption{Generalized relaxation and decoherence times $T_{1}$ and $T_{2}$
for the transverse field Ising chains of different lengths $L$ with
longitudinal coupling as functions of the transverse field. The tip
is unpolarized, and we use the parameters $J=1$ meV, $g=1$, $k_{B}T=0.1$
meV, $J_{s}=J_{t}=1$ meV, and $W=10$ meV. The maxima of the curves
correspond to the level crossing points where the energies of the
qubit states coincide. }
\label{fgr:TFLCIsingH}
\end{figure}
\begin{figure}[t]

\includegraphics[width=1\columnwidth]{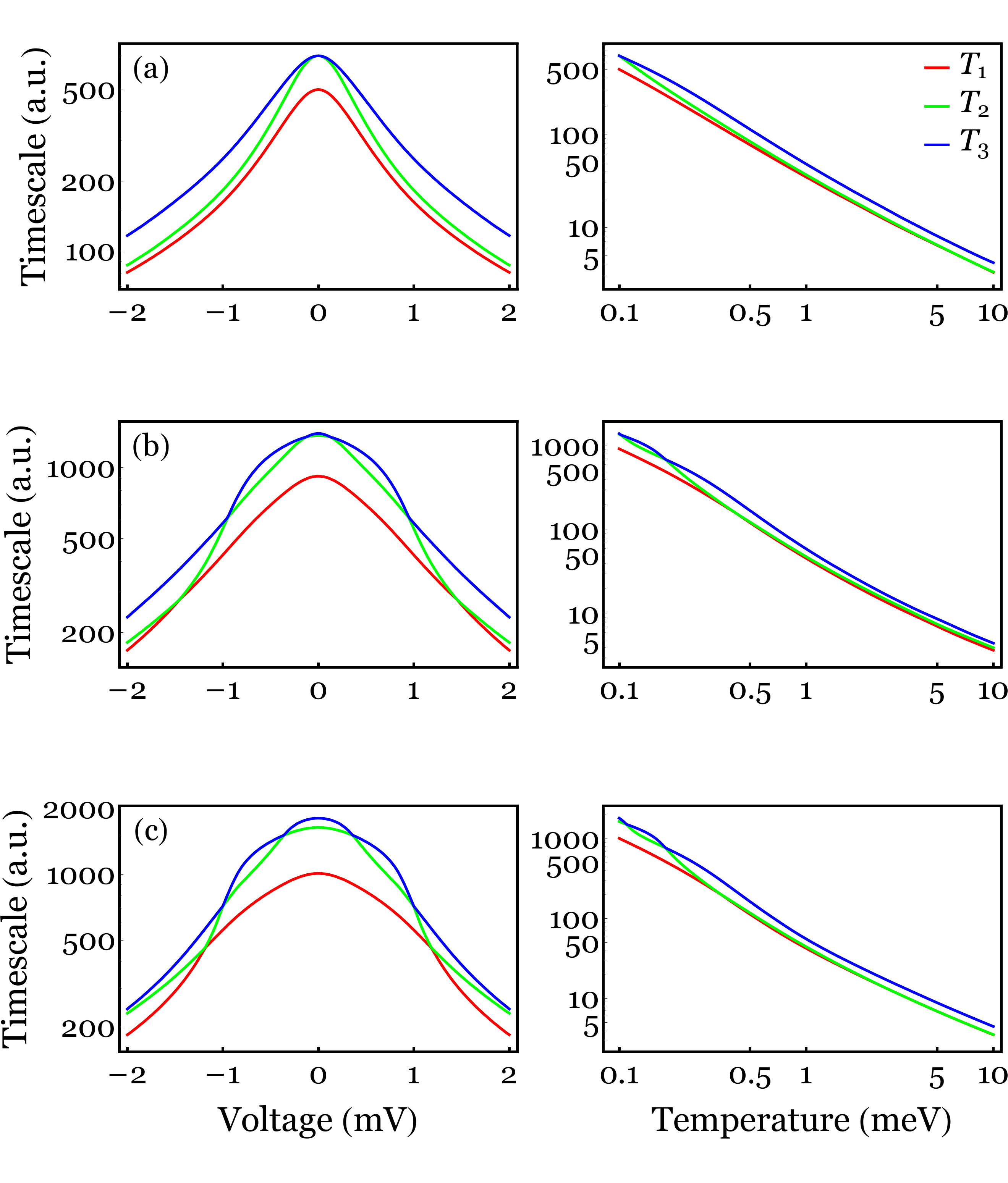}\caption{Timescales of information loss for the transverse field Ising chains
of different lengths with longitudinal coupling as functions of the
voltage and the temperature. The tip is unpolarized, and we use the
parameters $J=1$ meV, $g=1$, $J_{s}=J_{t}=1$ meV, and $W=10$ meV.
The chain lengths are (a) $L=2$, (b) $L=3$, (c) $L=4$. The values
of $h$ are different for each $L$ and correspond to the largest
level crossing points. For the voltage curves the temperature is fixed
at $k_{B}T=0.1$ meV; for the temperature curves the voltage is fixed
at zero.}
\label{fgr:TFLCIsingVT}
\end{figure}

Finally, we study the properties of the relaxation and decoherence
times when topological states are used for qubit realization. For
this we consider the transverse field Ising chain with longitudinal
coupling described by the Hamiltonian
\begin{align}
H_{S} & =J\sum_{l=1}^{L-1}\left(S_{lx}S_{l+1,x}+gS_{lz}S_{l+1,z}\right)-h\sum_{l=1}^{L}S_{lz},
\end{align}
where we assume the parameters $J$, $g$, and $h$ to have positive
values. In the thermodynamic limit $L\to\infty$ this model experiences
a transition from paramagnetic to magnetically ordered phase upon
tuning the field value $h$. In the paramagnetic phase, below the
critical value of the field, the system has a twofold degenerate ground
state which might be described in terms of the Majorana edge states
\cite{Kitaev2001}. For finite chains the degeneracy is lifted but
the topological order manifests itself in the level crossings of the
two lowest energy states as the field value is changed \cite{Toskovic2016}.
Figure \ref{fgr:TFLCIsingH} shows the dependence of the times $T_{1}$
and $T_{2}$ for the qubit realized with these states on the field
value $h$ for the chains of different length $L$. In the calculations
we assume the unpolarized tip and use the parameters $J=1$ meV, $g=1$,
$k_{B}T=0.1$ meV for the temperature, $J_{s}=J_{t}=1$ meV for the inelastic coupling to the leads, and $W=10$ meV for the bandwidth of the leads. One sees
an increase in the relaxation and decoherence times at values of $h$
corresponding to the level crossings. This behavior can be explained
by the dependence of the transition rates on the excitation energy,
see Appendix \ref{sec:RateEquations}, and has nothing to do with
the topological nature of the qubit states. Thus the topological order
does not suppress environment induced decoherence in this case. If
the field value is kept at its critical value, three timescales of
the information loss associated with the decay of the quantity (\ref{eq:normdistance})
do not contain a pair of close values, as was the case in all previous
examples. The dependence of all three timescales on the voltage and
the temperature is shown in Fig. \ref{fgr:TFLCIsingVT}, where the
value of the field is chosen at the largest level crossing point for
each $L$. One might see that $T_{1}$, $T_{2}$, and $T_{3}$ are
of the same order and demonstrate similar behavior. At low temperatures,
these timescales do not show an exponential growth because direct
transitions between $\ket 0$ and $\ket 1$ are allowed, i.e., $\bra 0S_{la}\ket 1\neq0$
in general.

\section{Conclusion \label{sec:Conclusion}}

In conclusion, we applied a master equation approach to study relaxation
and decoherence in engineered atomic spin structures due to their
interaction with electrons from surroundings. Our goal was to characterize
these two processes by the corresponding timescales $T_{1}$ and $T_{2}$
for a qubit realized on the two lowest energy states of an atomic
spin structure. We thus analyzed the dynamics of the corresponding
$2\times2$ subspace of the atomic density matrix. We encountered
a problem with the usual association of the relaxation and decoherence
times with the decay of the population imbalance and the off-diagonal
matrix elements. The only case when the standard definition can be
used is for a single atom with spin $S=1/2$ whose density matrix
can be unequivocally mapped onto the Bloch vector. But even in this
case, the standard definition has limited applicability since it implicitly
assumes dynamics of coherences and populations to be decoupled. This
condition is not fulfilled if the tip polarization is not parallel
to the polarization of the qubit states. For systems with more than
two states, the problem of defining $T_{1}$ and $T_{2}$ gets even
more severe, as the dynamics of the populations and coherences within
the qubit subspace couple to higher energy states. In the present
paper, we generalize the concepts of the relaxation and
decoherence times by introducing three timescales $T_{1}$, $T_{2}$, and $T_{3}$ that characterize the process of the information loss. These timescales match the standard relaxation and decoherence times, with $T_{2}=T_{3}$, whenever the latter two are well defined.

Equipped with the generalized definitions for the relaxation and decoherence
times, we studied several classes of systems where the qubit states
are identified with the two lowest energy states. We found that in most cases $T_{2}\simeq T_{3}$, and one can reduce three timescales to two, associating $T_{1}$ with relaxation and $T_{2,3}$ with decoherence.
Analysis of single atoms of different spin and spin chains of different
length showed that $T_{1}$ generically increases with the system
size, while $T_{2}$ does the opposite. The growth of $T_{1}$ with
system size is faster for the collective spin models where not only
the number of cascade transitions, needed to pass from one low-lying
energy state to another, increases, but the energies of the transitions
also do. We found that both relaxation and decoherence times behave
as $1/T$ and $1/V$ in the limits of large temperature and large
voltage correspondingly. The low temperature behavior of $T_{1}$
depends on whether the environment is able to induce direct transitions
between the qubit states. If not, the relaxation time grows exponentially
as the temperature is decreased, due to the Boltzmann weight. Strategies
to increase $T_{1}$ pass therefore by providing an energy barrier
for transitions between the two states. The low temperature properties
of $T_{2}$ are determined by the average polarizations of the atoms
in the qubit states. If they are not zero, the environment causes
dephasing and $T_{2}\propto1/T$ behavior is preserved. If all the
atoms are unpolarized in both qubit states, the dephasing is absent
and the decoherence is fully due to transitions to higher energy states,
i.e., $T_{2}$ behaves similarly to $T_{1}$ and might be exponentially
suppressed as well by lowering the temperature. This situation was
observed in the collective spin model where both qubit states are
singlets.

In more complex systems, here illustrated by a transverse field Ising
chain with longitudinal coupling tuned such that the ground state
is degenerate, the three timescales are seen, i.e., $T_{2}\neq T_{3}$.
In this case, the demarcation between the relaxation and decoherence processes is not possible since they are substantially coupled to each other.

This work gives specific directions towards the engineering of decoherence-free
spin based qubits and shows that the implementation of this strategy
in engineered atomic spin devices is realistic. We note that our model
includes only leading order processes in the atom-lead coupling and
assumes a decorated environment for each atom. At very low temperatures,
higher order or nonperturbative effects may limit the exponential
growth of $T_{2}$. 

The sources of decoherence are not only limited to the electrons in
the metallic surface, as the magnetic degrees of freedom also couple
to substrate phonons and nuclear spins \cite{Delgado2017}. The present
strategy will improve decoherence times as long as the leading coupling
to the environment involves a magnetic exchange interaction, as in
the case of nuclear spins. 

Our work motivates several questions. One is the possibility of implementing
a decoherence-free subspace for a single qubit whose $T_{1}$ and
$T_{2}$ can be made larger by increasing the number of spins in the
structure. This would permit us to improve the decoherence times arbitrarily
by increasing the resources allocated to encode each qubit. The use
of topologically protected states \cite{Dennis2002,Delgado2013,Nadj-Perge2014}
and skyrmions \cite{Romming2013,Nagaosa2013} indeed goes in this
direction, however it is hard to envision its implementation with
EASDs. Another question is how to scale this strategy to multiple
interacting qubits and how to perform current measurements that can
distinguish the qubit states. 
\begin{acknowledgments}
We are thankful to L.-H. Frahm for fruitful discussions. This work
was funded by RSF Grant No. 16-42-01057 and DFG Grant No. LI 1413/9-1.
P.R. acknowledges support by FCT through the Investigador FCT contract
IF/00347/2014 and Grant No. UID/CTM/04540/2013.
\end{acknowledgments}

\appendix

\section{Rate equations\label{sec:RateEquations}}

The rate equations for the populations of spin states
\begin{align}
\partial_{t}p_{\alpha} & =\sum_{\beta}R_{\alpha\beta}p_{\beta}-R_{\beta\alpha}p_{\alpha}\label{eq:REequation}
\end{align}
are obtained by substituting the density matrix of the form $\rho_{\alpha\beta}=p_{\alpha}\delta_{\alpha\beta}$
into the Redfield equation (\ref{eq:redfield}). The transition rates
in Eq. (\ref{eq:REequation}) are given by
\begin{align}
R_{\alpha\beta} & =2\re\sum_{laa'}\bra{\alpha}\varLambda_{laa'}\ket{\beta}\bra{\beta}S_{la}\ket{\alpha}\label{eq:RErates}
\end{align}
and might be expressed as
\begin{align}
R_{\alpha\beta} & =\frac{2}{\beta}\re\sum_{laa'}\bra{\alpha}S_{la'}\ket{\beta}\bra{\beta}S_{la}\ket{\alpha}\\
 & \times\sum_{\nu\nu'}u_{laa'}^{\nu\nu'}g\left(\beta\left(\omega_{\alpha}-\omega_{\beta}-\mu_{\nu}+\mu_{\nu'}\right)\right),\nonumber 
\end{align}
where $g\left(x\right)=x/\left(e^{x}-1\right)$. Note that this expression
does not contain the imaginary part of $\kappa\left(\omega\right)$
defined in Eq. (\ref{eq:kfunction}).

If the tip is absent, the transition rates are given by
\begin{align}
R_{\alpha\beta} & =\frac{\varGamma_{s}^{2}}{\pi\beta}\,g\left(\beta\left(\omega_{\alpha}-\omega_{\beta}\right)\right)\sum_{la}\left|\bra{\alpha}S_{la}\ket{\beta}\right|^{2}.
\end{align}
In the low temperature case, when transitions from the qubit states
$\ket 0$ and $\ket 1$ to the higher energy states are exponentially
suppressed, the relaxation time might be approximated by
\begin{align}
\frac{1}{T_{1}} & =R_{01}+R_{10}\label{eq:RErelaxtime}\\
 & =\frac{\varGamma_{s}^{2}}{\pi\beta}\left[g\left(\beta\Delta\right)+g\left(-\beta\Delta\right)\right]\sum_{la}\left|\bra 0S_{la}\ket 1\right|^{2},\nonumber 
\end{align}
where $\Delta=\omega_{1}-\omega_{0}$ is the energy splitting between
the qubit states. The function $T_{1}\left(\Delta\right)$ reaches
its maximum at $\Delta=0$ which explains Fig. \ref{fgr:TFLCIsingH}
discussed in Sec. \ref{subsec:Examples}. One may also see that the
relaxation time obtained from Eq. (\ref{eq:RErelaxtime}) is infinite
if the matrix elements of all local spin operators $S_{la}$ between
qubit states are zero, i.e., $\bra 0S_{la}\ket 1=0$. The real value
of $T_{1}$ in this case is determined by transition rates to higher
energy states which are exponentially small.

\section{Comparison of master equations\label{sec:Comparison}}

\begin{figure}[t]
\includegraphics[width=1\columnwidth]{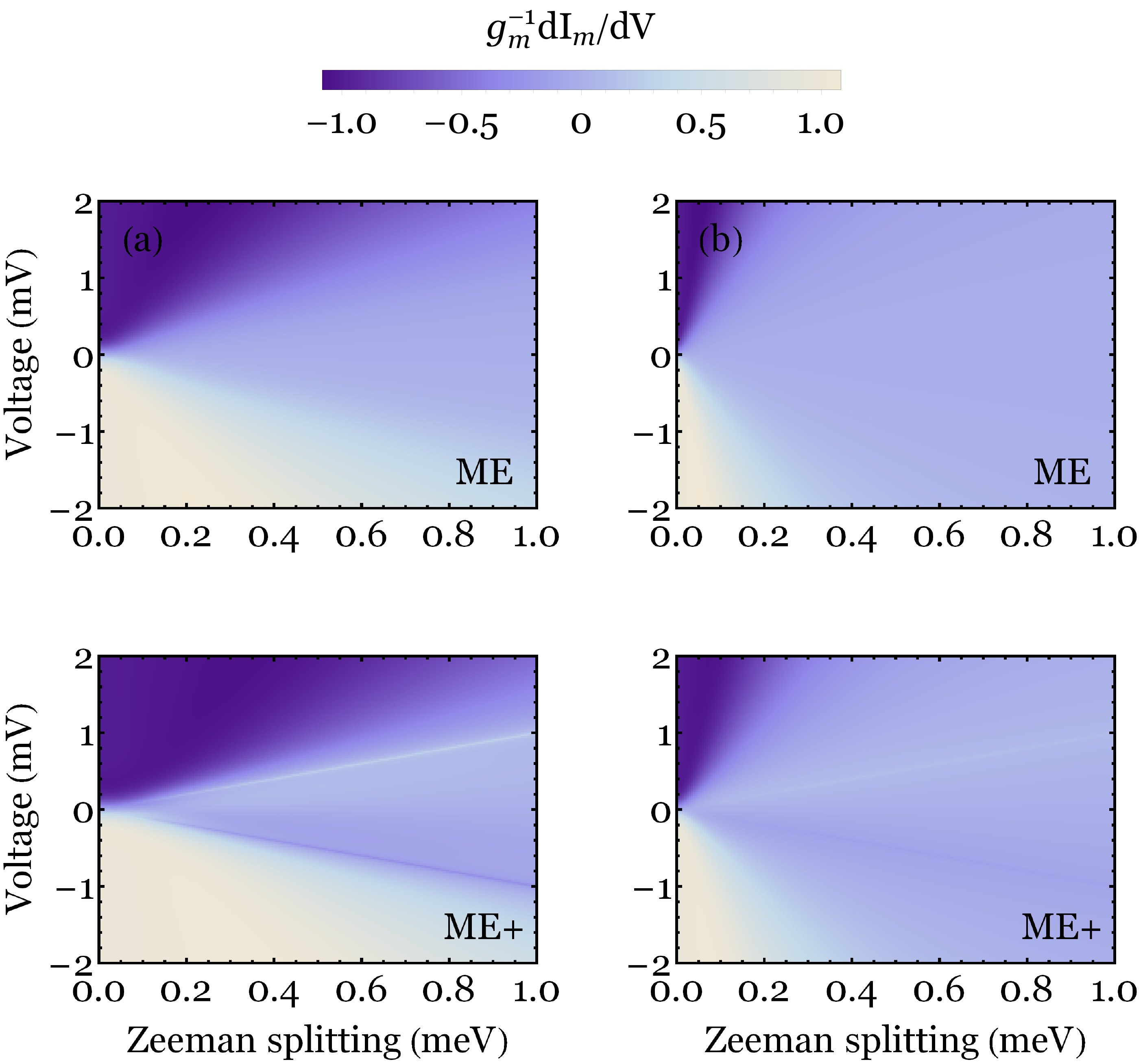}

\caption{Magnetoresistive contribution to the differential conductance for
a single atom with spin $S=1/2$ in the perpendicular magnetic field
($\boldsymbol{B}$ along $z$ axis, $\boldsymbol{P}$ along $x$ axis)
for different values of the bandwidth: (a) $W=10$ meV, (b) $W=20$
meV. The spectra are presented as functions of the voltage and the
Zeeman splitting and calculated with (i) master equation with only
real part of $\kappa\left(\omega\right)$ (ME), and (ii) master equation
with full $\kappa\left(\omega\right)$ (ME+). The rate equations give
zero for $dI_{m}/dV$. We use the parameters $k_{B}T=0.1$ meV, $J_{t}=J_{s}=5$
meV, and $p=1$.}

\label{fgr:mag_compare}
\end{figure}

\begin{figure}[t]
\includegraphics[width=1\columnwidth]{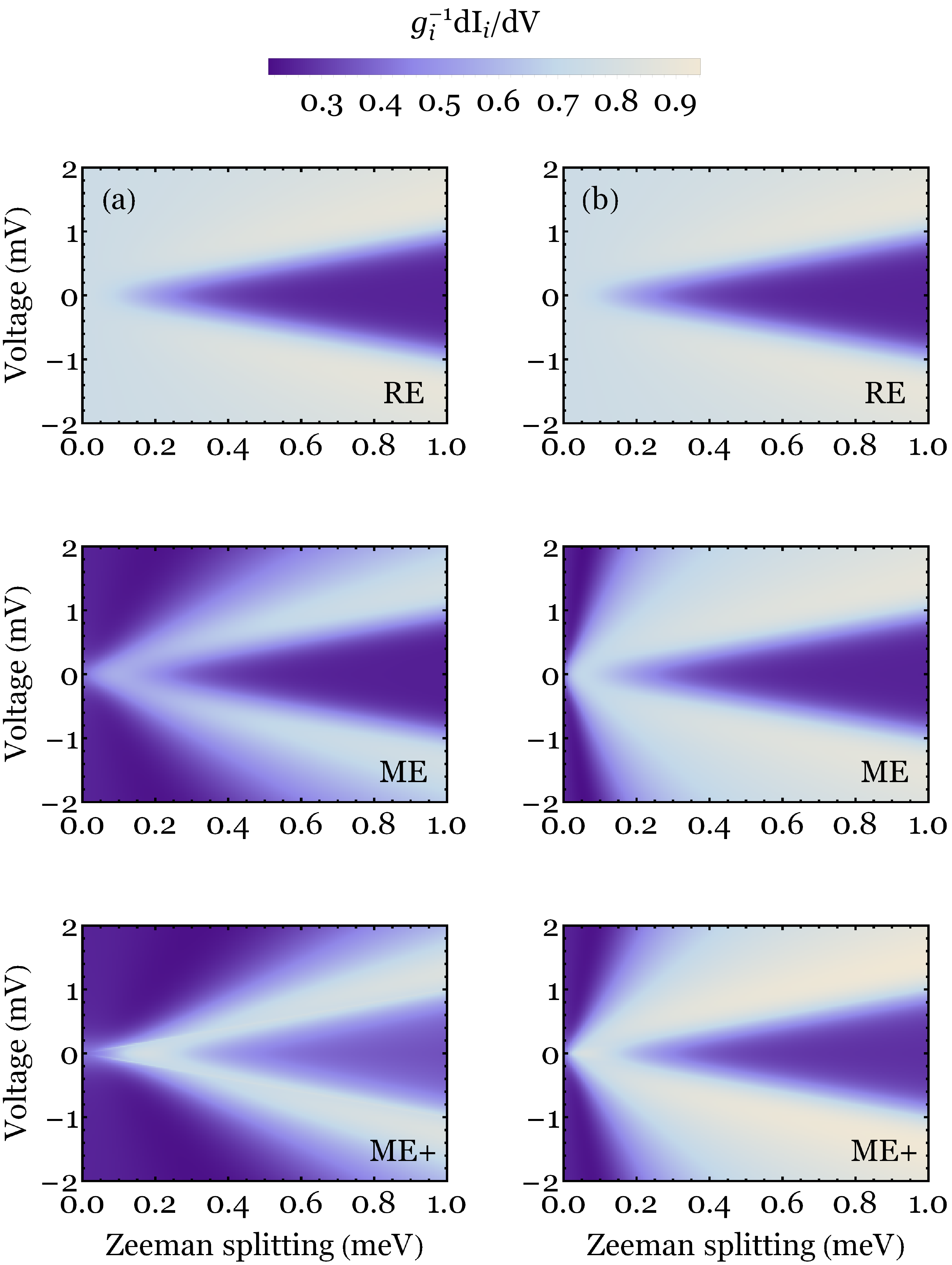}

\caption{Inelastic contribution to the differential conductance for a single
atom with spin $S=1/2$ in the perpendicular magnetic field ($\boldsymbol{B}$
along $z$ axis, $\boldsymbol{P}$ along $x$ axis) for different
values of the bandwidth: (a) $W=10$ meV, (b) $W=20$ meV. The spectra
are presented as functions of the voltage and the Zeeman splitting
and are calculated with (i) rate equations (RE), (ii) master equation
with only real part of $\kappa\left(\omega\right)$ (ME), and (iii)
master equation with full $\kappa\left(\omega\right)$ (ME+). We use
the parameters $k_{B}T=0.1$ meV, $J_{t}=J_{s}=5$ meV, and $p=1$.}

\label{fgr:inel_compare}
\end{figure}

In this appendix we compare the results for the magnetoresistive and
inelastic current spectra calculated with (i) rate equations, (ii)
master equation with only real part of $\kappa\left(\omega\right)$,
(iii) master equation with full $\kappa\left(\omega\right)$. We consider
a single atom with spin $S=1/2$ in the magnetic field perpendicular
to the polarization of the current, the simplest example where all
three methods give different results.

The spectra are presented in Figs. \ref{fgr:mag_compare} and \ref{fgr:inel_compare}
where their dependence on the Zeeman splitting is shown. In the calculation
we use the parameters $p=1$, $k_{B}T=0.1$ meV, $J_{t}=J_{s}=5$
meV, and vary the bandwidth $W=10,\,20$ meV. One might see that the
results calculated with three methods are different from each other.
However, the comparison of the plots for different bandwidth indicates
the decrease of the discrepancy as the coupling gets smaller. In the
large $W$ limit, the results of both master equation methods approach
the results of the RE method. This behavior is explained by the vanishing
of the spin transfer torque that is responsible for sustaining coherences
in the steady state \cite{Shakirov2016a}. In the opposite case, when
$W$ is decreased, the effect of coherences becomes more pronounced,
and the terms originating from the imaginary part of $\kappa\left(\omega\right)$
come into play. However, for the bandwidths smaller than some critical
value $W_{c}$, the steady state density matrix becomes unphysical.
This shortcoming is due to the limited applicability of the master
equation approach to open quantum systems strongly coupled to the
surroundings.

\section{Bloch equation for $S=1/2$\label{sec:BlochEquation}}

To map Eq. (\ref{eq:redfield}) onto the Bloch equation for a single
atom with spin $S=1/2$, we use the relations (\ref{eq:ptorho}) and
(\ref{eq:rhotop}) connecting the density matrix and the Bloch vector.
One obtains
\begin{align}
\partial_{t}p_{a} & =\tr\left(\sigma^{a}\partial_{t}\rho\right)=-i\tr\left(\left[\sigma^{a},H_{S}\right]\rho\right)\label{eq:maptobloch}\\
 & +\frac{1}{2}\sum_{a'}p_{a'}\sum_{b}\tr\left(\sigma^{a}\left(\left[\Lambda_{b}\sigma^{a'},S_{b}\right]+\textrm{H.c.}\right)\right)\nonumber \\
 & +\frac{1}{2}\sum_{b}\tr\left\{ \sigma^{a}\left(\left[\Lambda_{b},S_{b}\right]+\textrm{H.c.}\right)\right\} ,\,a=x,y,z,\nonumber 
\end{align}
where all matrices are taken in the eigenbasis. With $H_{S}=-\frac{1}{2}h\sigma^{z}$,
the unitary part of the Redfield equation becomes
\begin{align}
\frac{ih}{2}\tr\left(\left[\sigma^{a},\sigma^{z}\right]\rho\right) & =-h\sum_{b}\varepsilon_{azb}\tr\left(\sigma^{b}\rho\right)\\
 & =-h\left(\hat{z}\times\boldsymbol{p}\right)_{a},\nonumber 
\end{align}
and Eqs. (\ref{eq:maptobloch}) may be rewritten in the form
\begin{equation}
\partial_{t}\boldsymbol{p}=-\boldsymbol{\omega}_{0}\times\boldsymbol{p}-\boldsymbol{R}_{0}.\boldsymbol{p}+\boldsymbol{F},
\end{equation}
where $\boldsymbol{\omega}_{0}=h\hat{z}$, and we defined
\begin{align}
\left(R_{0}\right)_{aa'} & =-\frac{1}{2}\tr\left\{ \sigma^{a}\sum_{b}\left(\left[\Lambda_{b}\sigma^{a'},S_{b}\right]+\textrm{H.c.}\right)\right\} ,\label{eq:Rmatrix}\\
F_{a} & =\frac{1}{2}\tr\left\{ \sigma^{a}\sum_{b}\left(\left[\Lambda_{b},S_{b}\right]+\textrm{H.c.}\right)\right\} .\label{eq:Fvector}
\end{align}
We split the matrix $\boldsymbol{R}_{0}$ into the symmetric part
$\left(\boldsymbol{R}_{0}+\boldsymbol{R}_{0}^{T}\right)/2=\boldsymbol{R}$
and the antisymmetric one $\left(\boldsymbol{R}_{0}-\boldsymbol{R}_{0}^{T}\right)/2=\Delta\boldsymbol{\omega}\times$
to obtain the Bloch equation (\ref{eq:Bloch}), where $\boldsymbol{\omega}=\boldsymbol{\omega}_{0}+\Delta\boldsymbol{\omega}$
and $\boldsymbol{p}_{0}=\left(\boldsymbol{\omega}\times+\boldsymbol{R}\right)^{-1}.\boldsymbol{F}$.

\section{Generalized timescales of relaxation and decoherence for $S=1/2$\label{sec:TimesForTLS}}

In this appendix we apply the definitions for the generalized timescales of relaxation and decoherence introduced in Sec. \ref{subsec:Generalized-relaxation-and} to the case of a single atom with spin $S=1/2$ and compare them with the standard relaxation and decoherence times $T_{1}$ and $T_{2}$ from Eq. (\ref{eq:Rmatrixdecoupled}).

For an unpolarized tip, when dynamics of the populations and coherences are decoupled, the superoperator of the master equation has the matrix form
\begin{equation}
\mathcal{L}=
\left(
\begin{array}{cccc}
-\frac{1-p_{0}}{2T_{1}} & 0 & 0 & \frac{1+p_{0}}{2T_{1}} \\
0 & -\frac{1}{T_{2}}+i\omega & 0 & 0 \\
0 & 0 & -\frac{1}{T_{2}}-i\omega & 0 \\
\frac{1-p_{0}}{2T_{1}} & 0 & 0 & -\frac{1+p_{0}}{2T_{1}}
\end{array}
\right).
\label{eq:LsuperTLS}
\end{equation}
We obtained it from the Bloch equation (\ref{eq:Bloch}) using the relations (\ref{eq:ptorho}), (\ref{eq:rhotop}), (\ref{eq:Rmatrixdecoupled}), and the fact that $\boldsymbol{p}_{0}$ and $\boldsymbol{\omega}$ are aligned with $\hat{z}$. The right and left eigenvectors of the superoperator $\mathcal{L}$ are
\begin{equation}
\begin{split}
& \kket{\rho_{0}}=\left(
\begin{array}{cc}
\frac{1+p_{0}}{2} & 0 \\
0 & \frac{1-p_{0}}{2}
\end{array}
\right),~~~\kket{\tilde{\rho}_{0}}=\left(
\begin{array}{cc}
1 & 0 \\
0 & 1
\end{array}
\right), \\
& \kket{\rho_{1}}=\left(
\begin{array}{cc}
1 & 0 \\
0 & -1
\end{array}
\right),~~~\kket{\tilde{\rho}_{1}}=\left(
\begin{array}{cc}
\frac{1-p_{0}}{2} & 0 \\
0 & -\frac{1+p_{0}}{2}
\end{array}
\right), \\
& \kket{\rho_{2}}=\kket{\tilde{\rho}_{2}}=\left(
\begin{array}{cc}
0 & 1 \\
0 & 0
\end{array}
\right), \\
& \kket{\rho_{3}}=\kket{\tilde{\rho}_{3}}=\left(
\begin{array}{cc}
0 & 0 \\
1 & 0
\end{array}
\right),
\end{split}
\label{eq:RhoRL}
\end{equation}
with the eigenvalues given by $\lambda_{0}=0$ and Eqs. (\ref{eq:T1definition}) and (\ref{eq:T2definition}). Substitution of the eigenvectors (\ref{eq:RhoRL}) into the matrix elements (\ref{eq:Mmatrix}) results in
\begin{equation}
\boldsymbol{M}\left(t\right)=\left(
\begin{array}{ccc}
e^{-2t/T_{2}} & 0 & 0 \\
0 & e^{-2t/T_{2}} & 0 \\
0 & 0 & e^{-2t/T_{1}}
\end{array}
\right).
\end{equation}
One can see that two of three decay times of the eigenvalues of $\boldsymbol{M}\left(t\right)$ coincide with $T_{2}$, and the third is equal to $T_{1}$. Thus the definitions introduced in Sec. \ref{subsec:Generalized-relaxation-and} reproduce the standard relaxation and decoherence timescales for two-level systems.\\

\bibliographystyle{apsrev4-1}
\bibliography{RelaxBib}

\begin{thebibliography}{63}%
\makeatletter
\providecommand \@ifxundefined [1]{%
 \@ifx{#1\undefined}
}%
\providecommand \@ifnum [1]{%
 \ifnum #1\expandafter \@firstoftwo
 \else \expandafter \@secondoftwo
 \fi
}%
\providecommand \@ifx [1]{%
 \ifx #1\expandafter \@firstoftwo
 \else \expandafter \@secondoftwo
 \fi
}%
\providecommand \natexlab [1]{#1}%
\providecommand \enquote  [1]{``#1''}%
\providecommand \bibnamefont  [1]{#1}%
\providecommand \bibfnamefont [1]{#1}%
\providecommand \citenamefont [1]{#1}%
\providecommand \href@noop [0]{\@secondoftwo}%
\providecommand \href [0]{\begingroup \@sanitize@url \@href}%
\providecommand \@href[1]{\@@startlink{#1}\@@href}%
\providecommand \@@href[1]{\endgroup#1\@@endlink}%
\providecommand \@sanitize@url [0]{\catcode `\\12\catcode `\$12\catcode
  `\&12\catcode `\#12\catcode `\^12\catcode `\_12\catcode `\%12\relax}%
\providecommand \@@startlink[1]{}%
\providecommand \@@endlink[0]{}%
\providecommand \url  [0]{\begingroup\@sanitize@url \@url }%
\providecommand \@url [1]{\endgroup\@href {#1}{\urlprefix }}%
\providecommand \urlprefix  [0]{URL }%
\providecommand \Eprint [0]{\href }%
\providecommand \doibase [0]{http://dx.doi.org/}%
\providecommand \selectlanguage [0]{\@gobble}%
\providecommand \bibinfo  [0]{\@secondoftwo}%
\providecommand \bibfield  [0]{\@secondoftwo}%
\providecommand \translation [1]{[#1]}%
\providecommand \BibitemOpen [0]{}%
\providecommand \bibitemStop [0]{}%
\providecommand \bibitemNoStop [0]{.\EOS\space}%
\providecommand \EOS [0]{\spacefactor3000\relax}%
\providecommand \BibitemShut  [1]{\csname bibitem#1\endcsname}%
\let\auto@bib@innerbib\@empty
%</preamble>
\bibitem [{\citenamefont {Heinrich}\ \emph {et~al.}(2004)\citenamefont
  {Heinrich}, \citenamefont {Gupta}, \citenamefont {Lutz},\ and\ \citenamefont
  {Eigler}}]{Heinrich2004}%
  \BibitemOpen
  \bibfield  {author} {\bibinfo {author} {\bibfnamefont {A.~J.}\ \bibnamefont
  {Heinrich}}, \bibinfo {author} {\bibfnamefont {J.~A.}\ \bibnamefont {Gupta}},
  \bibinfo {author} {\bibfnamefont {C.~P.}\ \bibnamefont {Lutz}}, \ and\
  \bibinfo {author} {\bibfnamefont {D.~M.}\ \bibnamefont {Eigler}},\ }\href
  {\doibase 10.1126/science.1101077} {\bibfield  {journal} {\bibinfo  {journal}
  {Science}\ }\textbf {\bibinfo {volume} {306}},\ \bibinfo {pages} {466}
  (\bibinfo {year} {2004})}\BibitemShut {NoStop}%
\bibitem [{\citenamefont {Hirjibehedin}\ \emph {et~al.}(2007)\citenamefont
  {Hirjibehedin}, \citenamefont {Lin}, \citenamefont {Otte}, \citenamefont
  {Ternes}, \citenamefont {Lutz}, \citenamefont {Jones},\ and\ \citenamefont
  {Heinrich}}]{Hirjibehedin2007}%
  \BibitemOpen
  \bibfield  {author} {\bibinfo {author} {\bibfnamefont {C.~F.}\ \bibnamefont
  {Hirjibehedin}}, \bibinfo {author} {\bibfnamefont {C.-Y.}\ \bibnamefont
  {Lin}}, \bibinfo {author} {\bibfnamefont {A.~F.}\ \bibnamefont {Otte}},
  \bibinfo {author} {\bibfnamefont {M.}~\bibnamefont {Ternes}}, \bibinfo
  {author} {\bibfnamefont {C.~P.}\ \bibnamefont {Lutz}}, \bibinfo {author}
  {\bibfnamefont {B.~A.}\ \bibnamefont {Jones}}, \ and\ \bibinfo {author}
  {\bibfnamefont {A.~J.}\ \bibnamefont {Heinrich}},\ }\href {\doibase
  10.1126/science.1146110} {\bibfield  {journal} {\bibinfo  {journal}
  {Science}\ }\textbf {\bibinfo {volume} {317}},\ \bibinfo {pages} {1199}
  (\bibinfo {year} {2007})}\BibitemShut {NoStop}%
\bibitem [{\citenamefont {Tsukahara}\ \emph {et~al.}(2009)\citenamefont
  {Tsukahara}, \citenamefont {Noto}, \citenamefont {Ohara}, \citenamefont
  {Shiraki}, \citenamefont {Takagi}, \citenamefont {Takata}, \citenamefont
  {Miyawaki}, \citenamefont {Taguchi}, \citenamefont {Chainani}, \citenamefont
  {Shin},\ and\ \citenamefont {Kawai}}]{Tsukahara2009}%
  \BibitemOpen
  \bibfield  {author} {\bibinfo {author} {\bibfnamefont {N.}~\bibnamefont
  {Tsukahara}}, \bibinfo {author} {\bibfnamefont {K.~I.}\ \bibnamefont {Noto}},
  \bibinfo {author} {\bibfnamefont {M.}~\bibnamefont {Ohara}}, \bibinfo
  {author} {\bibfnamefont {S.}~\bibnamefont {Shiraki}}, \bibinfo {author}
  {\bibfnamefont {N.}~\bibnamefont {Takagi}}, \bibinfo {author} {\bibfnamefont
  {Y.}~\bibnamefont {Takata}}, \bibinfo {author} {\bibfnamefont
  {J.}~\bibnamefont {Miyawaki}}, \bibinfo {author} {\bibfnamefont
  {M.}~\bibnamefont {Taguchi}}, \bibinfo {author} {\bibfnamefont
  {A.}~\bibnamefont {Chainani}}, \bibinfo {author} {\bibfnamefont
  {S.}~\bibnamefont {Shin}}, \ and\ \bibinfo {author} {\bibfnamefont
  {M.}~\bibnamefont {Kawai}},\ }\href {\doibase 10.1103/PhysRevLett.102.167203}
  {\bibfield  {journal} {\bibinfo  {journal} {Phys. Rev. Lett.}\ }\textbf
  {\bibinfo {volume} {102}},\ \bibinfo {pages} {167203} (\bibinfo {year}
  {2009})}\BibitemShut {NoStop}%
\bibitem [{\citenamefont {Gauyacq}\ \emph {et~al.}(2012)\citenamefont
  {Gauyacq}, \citenamefont {Lorente},\ and\ \citenamefont
  {Novaes}}]{Gauyacq2012}%
  \BibitemOpen
  \bibfield  {author} {\bibinfo {author} {\bibfnamefont {J.-P.}\ \bibnamefont
  {Gauyacq}}, \bibinfo {author} {\bibfnamefont {N.}~\bibnamefont {Lorente}}, \
  and\ \bibinfo {author} {\bibfnamefont {F.~D.}\ \bibnamefont {Novaes}},\
  }\href {\doibase 10.1016/j.progsurf.2012.05.003} {\bibfield  {journal}
  {\bibinfo  {journal} {Progr. Surf. Sci.}\ }\textbf {\bibinfo {volume} {87}},\
  \bibinfo {pages} {63} (\bibinfo {year} {2012})}\BibitemShut {NoStop}%
\bibitem [{\citenamefont {Delgado}\ and\ \citenamefont
  {Fern{\'{a}}ndez-Rossier}(2010)}]{Delgado2010a}%
  \BibitemOpen
  \bibfield  {author} {\bibinfo {author} {\bibfnamefont {F.}~\bibnamefont
  {Delgado}}\ and\ \bibinfo {author} {\bibfnamefont {J.}~\bibnamefont
  {Fern{\'{a}}ndez-Rossier}},\ }\href {\doibase 10.1103/PhysRevB.82.134414}
  {\bibfield  {journal} {\bibinfo  {journal} {Phys. Rev. B}\ }\textbf {\bibinfo
  {volume} {82}},\ \bibinfo {pages} {134414} (\bibinfo {year}
  {2010})}\BibitemShut {NoStop}%
\bibitem [{\citenamefont {Delgado}\ \emph {et~al.}(2014)\citenamefont
  {Delgado}, \citenamefont {Hirjibehedin},\ and\ \citenamefont
  {Fernandez-Rossier}}]{Delgado2014}%
  \BibitemOpen
  \bibfield  {author} {\bibinfo {author} {\bibfnamefont {F.}~\bibnamefont
  {Delgado}}, \bibinfo {author} {\bibfnamefont {C.~F.}\ \bibnamefont
  {Hirjibehedin}}, \ and\ \bibinfo {author} {\bibfnamefont {J.}~\bibnamefont
  {Fernandez-Rossier}},\ }\href {\doibase 10.1016/j.susc.2014.07.009}
  {\bibfield  {journal} {\bibinfo  {journal} {Surf. Sci.}\ }\textbf {\bibinfo
  {volume} {630}},\ \bibinfo {pages} {337} (\bibinfo {year}
  {2014})}\BibitemShut {NoStop}%
\bibitem [{\citenamefont {Delgado}\ and\ \citenamefont
  {Fern{\'{a}}ndez-Rossier}(2017)}]{Delgado2017}%
  \BibitemOpen
  \bibfield  {author} {\bibinfo {author} {\bibfnamefont {F.}~\bibnamefont
  {Delgado}}\ and\ \bibinfo {author} {\bibfnamefont {J.}~\bibnamefont
  {Fern{\'{a}}ndez-Rossier}},\ }\href {\doibase 10.1103/PhysRevB.90.155134}
  {\bibfield  {journal} {\bibinfo  {journal} {Progr. Surf. Sci.}\ }\textbf
  {\bibinfo {volume} {92}},\ \bibinfo {pages} {40} (\bibinfo {year}
  {2017})}\BibitemShut {NoStop}%
\bibitem [{\citenamefont {Stipe}(1998)}]{Stipe1998}%
  \BibitemOpen
  \bibfield  {author} {\bibinfo {author} {\bibfnamefont {B.~C.}\ \bibnamefont
  {Stipe}},\ }\href {\doibase 10.1126/science.280.5370.1732} {\bibfield
  {journal} {\bibinfo  {journal} {Science}\ }\textbf {\bibinfo {volume}
  {280}},\ \bibinfo {pages} {1732} (\bibinfo {year} {1998})}\BibitemShut
  {NoStop}%
\bibitem [{\citenamefont {Meier}\ \emph {et~al.}(2008)\citenamefont {Meier},
  \citenamefont {Zhou}, \citenamefont {Wiebe},\ and\ \citenamefont
  {Wiesendanger}}]{Meier2008}%
  \BibitemOpen
  \bibfield  {author} {\bibinfo {author} {\bibfnamefont {F.}~\bibnamefont
  {Meier}}, \bibinfo {author} {\bibfnamefont {L.}~\bibnamefont {Zhou}},
  \bibinfo {author} {\bibfnamefont {J.}~\bibnamefont {Wiebe}}, \ and\ \bibinfo
  {author} {\bibfnamefont {R.}~\bibnamefont {Wiesendanger}},\ }\href {\doibase
  10.1126/science.1154415} {\bibfield  {journal} {\bibinfo  {journal}
  {Science}\ }\textbf {\bibinfo {volume} {320}},\ \bibinfo {pages} {82}
  (\bibinfo {year} {2008})}\BibitemShut {NoStop}%
\bibitem [{\citenamefont {Wiesendanger}(2009)}]{Wiesendanger2009}%
  \BibitemOpen
  \bibfield  {author} {\bibinfo {author} {\bibfnamefont {R.}~\bibnamefont
  {Wiesendanger}},\ }\href {\doibase 10.1103/RevModPhys.81.1495} {\bibfield
  {journal} {\bibinfo  {journal} {Rev. Mod. Phys.}\ }\textbf {\bibinfo {volume}
  {81}},\ \bibinfo {pages} {1495} (\bibinfo {year} {2009})}\BibitemShut
  {NoStop}%
\bibitem [{\citenamefont {Loth}\ \emph {et~al.}(2010)\citenamefont {Loth},
  \citenamefont {von Bergmann}, \citenamefont {Ternes}, \citenamefont {Otte},
  \citenamefont {Lutz},\ and\ \citenamefont {Heinrich}}]{Loth2010}%
  \BibitemOpen
  \bibfield  {author} {\bibinfo {author} {\bibfnamefont {S.}~\bibnamefont
  {Loth}}, \bibinfo {author} {\bibfnamefont {K.}~\bibnamefont {von Bergmann}},
  \bibinfo {author} {\bibfnamefont {M.}~\bibnamefont {Ternes}}, \bibinfo
  {author} {\bibfnamefont {A.~F.}\ \bibnamefont {Otte}}, \bibinfo {author}
  {\bibfnamefont {C.~P.}\ \bibnamefont {Lutz}}, \ and\ \bibinfo {author}
  {\bibfnamefont {A.~J.}\ \bibnamefont {Heinrich}},\ }\href {\doibase
  10.1038/nphys1616} {\bibfield  {journal} {\bibinfo  {journal} {Nat. Phys.}\
  }\textbf {\bibinfo {volume} {6}},\ \bibinfo {pages} {340} (\bibinfo {year}
  {2010})}\BibitemShut {NoStop}%
\bibitem [{\citenamefont {Baumann}\ \emph {et~al.}(2015)\citenamefont
  {Baumann}, \citenamefont {Paul}, \citenamefont {Choi}, \citenamefont {Lutz},
  \citenamefont {Ardavan},\ and\ \citenamefont {Heinrich}}]{Baumann2015}%
  \BibitemOpen
  \bibfield  {author} {\bibinfo {author} {\bibfnamefont {S.}~\bibnamefont
  {Baumann}}, \bibinfo {author} {\bibfnamefont {W.}~\bibnamefont {Paul}},
  \bibinfo {author} {\bibfnamefont {T.}~\bibnamefont {Choi}}, \bibinfo {author}
  {\bibfnamefont {C.~P.}\ \bibnamefont {Lutz}}, \bibinfo {author}
  {\bibfnamefont {A.}~\bibnamefont {Ardavan}}, \ and\ \bibinfo {author}
  {\bibfnamefont {A.~J.}\ \bibnamefont {Heinrich}},\ }\href {\doibase
  10.1126/science.aac8703} {\bibfield  {journal} {\bibinfo  {journal}
  {Science}\ }\textbf {\bibinfo {volume} {350}},\ \bibinfo {pages} {417}
  (\bibinfo {year} {2015})}\BibitemShut {NoStop}%
\bibitem [{\citenamefont {Krause}\ \emph {et~al.}(2016)\citenamefont {Krause},
  \citenamefont {Sonntag}, \citenamefont {Hermenau}, \citenamefont
  {Friedlein},\ and\ \citenamefont {Wiesendanger}}]{Krause2016}%
  \BibitemOpen
  \bibfield  {author} {\bibinfo {author} {\bibfnamefont {S.}~\bibnamefont
  {Krause}}, \bibinfo {author} {\bibfnamefont {A.}~\bibnamefont {Sonntag}},
  \bibinfo {author} {\bibfnamefont {J.}~\bibnamefont {Hermenau}}, \bibinfo
  {author} {\bibfnamefont {J.}~\bibnamefont {Friedlein}}, \ and\ \bibinfo
  {author} {\bibfnamefont {R.}~\bibnamefont {Wiesendanger}},\ }\href {\doibase
  10.1103/PhysRevB.93.064407} {\bibfield  {journal} {\bibinfo  {journal} {Phys.
  Rev. B}\ }\textbf {\bibinfo {volume} {93}},\ \bibinfo {pages} {064407}
  (\bibinfo {year} {2016})}\BibitemShut {NoStop}%
\bibitem [{\citenamefont {Paul}\ \emph {et~al.}(2016)\citenamefont {Paul},
  \citenamefont {Yang}, \citenamefont {Baumann}, \citenamefont {Romming},
  \citenamefont {Choi}, \citenamefont {Lutz},\ and\ \citenamefont
  {Heinrich}}]{Paul2016}%
  \BibitemOpen
  \bibfield  {author} {\bibinfo {author} {\bibfnamefont {W.}~\bibnamefont
  {Paul}}, \bibinfo {author} {\bibfnamefont {K.}~\bibnamefont {Yang}}, \bibinfo
  {author} {\bibfnamefont {S.}~\bibnamefont {Baumann}}, \bibinfo {author}
  {\bibfnamefont {N.}~\bibnamefont {Romming}}, \bibinfo {author} {\bibfnamefont
  {T.}~\bibnamefont {Choi}}, \bibinfo {author} {\bibfnamefont {C.~P.}\
  \bibnamefont {Lutz}}, \ and\ \bibinfo {author} {\bibfnamefont {A.~J.}\
  \bibnamefont {Heinrich}},\ }\href {\doibase 10.1038/nphys3965} {\bibfield
  {journal} {\bibinfo  {journal} {Nat. Phys.}\ }\textbf {\bibinfo {volume}
  {13}},\ \bibinfo {pages} {403} (\bibinfo {year} {2016})}\BibitemShut
  {NoStop}%
\bibitem [{\citenamefont {Lado}\ \emph {et~al.}(2016)\citenamefont {Lado},
  \citenamefont {Ferron},\ and\ \citenamefont {Fernandez-Rossier}}]{Lado2016}%
  \BibitemOpen
  \bibfield  {author} {\bibinfo {author} {\bibfnamefont {J.~L.}\ \bibnamefont
  {Lado}}, \bibinfo {author} {\bibfnamefont {A.}~\bibnamefont {Ferron}}, \ and\
  \bibinfo {author} {\bibfnamefont {J.}~\bibnamefont {Fernandez-Rossier}},\
  }\href
  {http://arxiv.org/abs/1611.01110{\%}0Ahttps://arxiv.org/abs/1611.01110}
  {\bibfield  {journal} {\bibinfo  {journal} {arXiv}\ } (\bibinfo {year}
  {2016})},\ \Eprint {http://arxiv.org/abs/1611.01110} {arXiv:1611.01110}
  \BibitemShut {NoStop}%
\bibitem [{\citenamefont {Yan}\ \emph {et~al.}(2017)\citenamefont {Yan},
  \citenamefont {Malavolti}, \citenamefont {Burgess}, \citenamefont
  {Droghetti}, \citenamefont {Rubio},\ and\ \citenamefont {Loth}}]{Yan2017}%
  \BibitemOpen
  \bibfield  {author} {\bibinfo {author} {\bibfnamefont {S.}~\bibnamefont
  {Yan}}, \bibinfo {author} {\bibfnamefont {L.}~\bibnamefont {Malavolti}},
  \bibinfo {author} {\bibfnamefont {J.~A.~J.}\ \bibnamefont {Burgess}},
  \bibinfo {author} {\bibfnamefont {A.}~\bibnamefont {Droghetti}}, \bibinfo
  {author} {\bibfnamefont {A.}~\bibnamefont {Rubio}}, \ and\ \bibinfo {author}
  {\bibfnamefont {S.}~\bibnamefont {Loth}},\ }\href {\doibase
  10.1126/sciadv.1603137} {\bibfield  {journal} {\bibinfo  {journal} {Sci.
  Adv.}\ }\textbf {\bibinfo {volume} {3}},\ \bibinfo {pages} {e1603137}
  (\bibinfo {year} {2017})}\BibitemShut {NoStop}%
\bibitem [{\citenamefont {Hirjibehedin}\ \emph {et~al.}(2006)\citenamefont
  {Hirjibehedin}, \citenamefont {Lutz},\ and\ \citenamefont
  {Heinrich}}]{Hirjibehedin2006}%
  \BibitemOpen
  \bibfield  {author} {\bibinfo {author} {\bibfnamefont {C.~F.}\ \bibnamefont
  {Hirjibehedin}}, \bibinfo {author} {\bibfnamefont {C.~P.}\ \bibnamefont
  {Lutz}}, \ and\ \bibinfo {author} {\bibfnamefont {A.~J.}\ \bibnamefont
  {Heinrich}},\ }\href {\doibase 10.1126/science.1125398} {\bibfield  {journal}
  {\bibinfo  {journal} {Science}\ }\textbf {\bibinfo {volume} {312}},\ \bibinfo
  {pages} {1021} (\bibinfo {year} {2006})}\BibitemShut {NoStop}%
\bibitem [{\citenamefont {Otte}\ \emph {et~al.}(2009)\citenamefont {Otte},
  \citenamefont {Ternes}, \citenamefont {Loth}, \citenamefont {Lutz},
  \citenamefont {Hirjibehedin},\ and\ \citenamefont {Heinrich}}]{Otte2009}%
  \BibitemOpen
  \bibfield  {author} {\bibinfo {author} {\bibfnamefont {A.~F.}\ \bibnamefont
  {Otte}}, \bibinfo {author} {\bibfnamefont {M.}~\bibnamefont {Ternes}},
  \bibinfo {author} {\bibfnamefont {S.}~\bibnamefont {Loth}}, \bibinfo {author}
  {\bibfnamefont {C.~P.}\ \bibnamefont {Lutz}}, \bibinfo {author}
  {\bibfnamefont {C.~F.}\ \bibnamefont {Hirjibehedin}}, \ and\ \bibinfo
  {author} {\bibfnamefont {A.~J.}\ \bibnamefont {Heinrich}},\ }\href {\doibase
  10.1103/PhysRevLett.103.107203} {\bibfield  {journal} {\bibinfo  {journal}
  {Phys. Rev. Lett.}\ }\textbf {\bibinfo {volume} {103}},\ \bibinfo {pages}
  {107203} (\bibinfo {year} {2009})}\BibitemShut {NoStop}%
\bibitem [{\citenamefont {Spinelli}\ \emph {et~al.}(2014)\citenamefont
  {Spinelli}, \citenamefont {Bryant}, \citenamefont {Delgado}, \citenamefont
  {Fern{\'{a}}ndez-Rossier},\ and\ \citenamefont {Otte}}]{Spinelli2014}%
  \BibitemOpen
  \bibfield  {author} {\bibinfo {author} {\bibfnamefont {A.}~\bibnamefont
  {Spinelli}}, \bibinfo {author} {\bibfnamefont {B.}~\bibnamefont {Bryant}},
  \bibinfo {author} {\bibfnamefont {F.}~\bibnamefont {Delgado}}, \bibinfo
  {author} {\bibfnamefont {J.}~\bibnamefont {Fern{\'{a}}ndez-Rossier}}, \ and\
  \bibinfo {author} {\bibfnamefont {A.~F.}\ \bibnamefont {Otte}},\ }\href
  {\doibase 10.1038/nmat4018} {\bibfield  {journal} {\bibinfo  {journal} {Nat.
  Mater.}\ }\textbf {\bibinfo {volume} {13}},\ \bibinfo {pages} {782} (\bibinfo
  {year} {2014})}\BibitemShut {NoStop}%
\bibitem [{\citenamefont {Choi}\ \emph {et~al.}(2016)\citenamefont {Choi},
  \citenamefont {Robles}, \citenamefont {Gauyacq}, \citenamefont {Ternes},
  \citenamefont {Loth},\ and\ \citenamefont {Lorente}}]{Choi2016}%
  \BibitemOpen
  \bibfield  {author} {\bibinfo {author} {\bibfnamefont {D.-J.}\ \bibnamefont
  {Choi}}, \bibinfo {author} {\bibfnamefont {R.}~\bibnamefont {Robles}},
  \bibinfo {author} {\bibfnamefont {J.-P.}\ \bibnamefont {Gauyacq}}, \bibinfo
  {author} {\bibfnamefont {M.}~\bibnamefont {Ternes}}, \bibinfo {author}
  {\bibfnamefont {S.}~\bibnamefont {Loth}}, \ and\ \bibinfo {author}
  {\bibfnamefont {N.}~\bibnamefont {Lorente}},\ }\href {\doibase
  10.1103/PhysRevB.94.085406} {\bibfield  {journal} {\bibinfo  {journal} {Phys.
  Rev. B}\ }\textbf {\bibinfo {volume} {94}},\ \bibinfo {pages} {085406}
  (\bibinfo {year} {2016})}\BibitemShut {NoStop}%
\bibitem [{\citenamefont {Girovsky}\ \emph {et~al.}(2017)\citenamefont
  {Girovsky}, \citenamefont {Lado}, \citenamefont {Kalff}, \citenamefont
  {Fahrenfort}, \citenamefont {Peters}, \citenamefont
  {Fern{\'{a}}ndez-Rossier},\ and\ \citenamefont {Otte}}]{Girovsky2017}%
  \BibitemOpen
  \bibfield  {author} {\bibinfo {author} {\bibfnamefont {J.}~\bibnamefont
  {Girovsky}}, \bibinfo {author} {\bibfnamefont {J.~L.}\ \bibnamefont {Lado}},
  \bibinfo {author} {\bibfnamefont {F.~E.}\ \bibnamefont {Kalff}}, \bibinfo
  {author} {\bibfnamefont {E.}~\bibnamefont {Fahrenfort}}, \bibinfo {author}
  {\bibfnamefont {L.~J. J.~M.}\ \bibnamefont {Peters}}, \bibinfo {author}
  {\bibfnamefont {J.}~\bibnamefont {Fern{\'{a}}ndez-Rossier}}, \ and\ \bibinfo
  {author} {\bibfnamefont {A.~F.}\ \bibnamefont {Otte}},\ }\href {\doibase
  10.21468/SciPostPhys.2.3.020} {\bibfield  {journal} {\bibinfo  {journal}
  {SciPost Phys.}\ }\textbf {\bibinfo {volume} {2}},\ \bibinfo {pages} {020}
  (\bibinfo {year} {2017})}\BibitemShut {NoStop}%
\bibitem [{\citenamefont {Imre}(2006)}]{Imre2006}%
  \BibitemOpen
  \bibfield  {author} {\bibinfo {author} {\bibfnamefont {A.}~\bibnamefont
  {Imre}},\ }\href {\doibase 10.1126/science.1120506} {\bibfield  {journal}
  {\bibinfo  {journal} {Science}\ }\textbf {\bibinfo {volume} {311}},\ \bibinfo
  {pages} {205} (\bibinfo {year} {2006})}\BibitemShut {NoStop}%
\bibitem [{\citenamefont {Khajetoorians}\ \emph {et~al.}(2011)\citenamefont
  {Khajetoorians}, \citenamefont {Wiebe}, \citenamefont {Chilian},\ and\
  \citenamefont {Wiesendanger}}]{Khajetoorians2011}%
  \BibitemOpen
  \bibfield  {author} {\bibinfo {author} {\bibfnamefont {A.~A.}\ \bibnamefont
  {Khajetoorians}}, \bibinfo {author} {\bibfnamefont {J.}~\bibnamefont
  {Wiebe}}, \bibinfo {author} {\bibfnamefont {B.}~\bibnamefont {Chilian}}, \
  and\ \bibinfo {author} {\bibfnamefont {R.}~\bibnamefont {Wiesendanger}},\
  }\href {\doibase 10.1126/science.1201725} {\bibfield  {journal} {\bibinfo
  {journal} {Science}\ }\textbf {\bibinfo {volume} {332}},\ \bibinfo {pages}
  {1062} (\bibinfo {year} {2011})}\BibitemShut {NoStop}%
\bibitem [{\citenamefont {Gauyacq}\ and\ \citenamefont
  {Lorente}(2015)}]{Gauyacq2015}%
  \BibitemOpen
  \bibfield  {author} {\bibinfo {author} {\bibfnamefont {J.-P.}\ \bibnamefont
  {Gauyacq}}\ and\ \bibinfo {author} {\bibfnamefont {N.}~\bibnamefont
  {Lorente}},\ }\href {\doibase 10.1088/0953-8984/27/45/455301} {\bibfield
  {journal} {\bibinfo  {journal} {J. Phys. Condens. Matter}\ }\textbf {\bibinfo
  {volume} {27}},\ \bibinfo {pages} {455301} (\bibinfo {year}
  {2015})}\BibitemShut {NoStop}%
\bibitem [{\citenamefont {Delgado}\ \emph {et~al.}(2015)\citenamefont
  {Delgado}, \citenamefont {Loth}, \citenamefont {Zielinski},\ and\
  \citenamefont {Fern{\'{a}}ndez-Rossier}}]{Delgado2015}%
  \BibitemOpen
  \bibfield  {author} {\bibinfo {author} {\bibfnamefont {F.}~\bibnamefont
  {Delgado}}, \bibinfo {author} {\bibfnamefont {S.}~\bibnamefont {Loth}},
  \bibinfo {author} {\bibfnamefont {M.}~\bibnamefont {Zielinski}}, \ and\
  \bibinfo {author} {\bibfnamefont {J.}~\bibnamefont
  {Fern{\'{a}}ndez-Rossier}},\ }\href {\doibase 10.1209/0295-5075/109/57001}
  {\bibfield  {journal} {\bibinfo  {journal} {Europhys. Lett.}\ }\textbf
  {\bibinfo {volume} {109}},\ \bibinfo {pages} {57001} (\bibinfo {year}
  {2015})}\BibitemShut {NoStop}%
\bibitem [{\citenamefont {Leuenberger}\ and\ \citenamefont
  {Loss}(2001)}]{Leuenberger2001}%
  \BibitemOpen
  \bibfield  {author} {\bibinfo {author} {\bibfnamefont {M.~N.}\ \bibnamefont
  {Leuenberger}}\ and\ \bibinfo {author} {\bibfnamefont {D.}~\bibnamefont
  {Loss}},\ }\href {\doibase 10.1038/35071024} {\bibfield  {journal} {\bibinfo
  {journal} {Nature (London)}\ }\textbf {\bibinfo {volume} {410}},\ \bibinfo
  {pages} {789} (\bibinfo {year} {2001})}\BibitemShut {NoStop}%
\bibitem [{\citenamefont {Troiani}\ \emph {et~al.}(2005)\citenamefont
  {Troiani}, \citenamefont {Ghirri}, \citenamefont {Affronte}, \citenamefont
  {Carretta}, \citenamefont {Santini}, \citenamefont {Amoretti}, \citenamefont
  {Piligkos}, \citenamefont {Timco},\ and\ \citenamefont
  {Winpenny}}]{Troiani2005}%
  \BibitemOpen
  \bibfield  {author} {\bibinfo {author} {\bibfnamefont {F.}~\bibnamefont
  {Troiani}}, \bibinfo {author} {\bibfnamefont {A.}~\bibnamefont {Ghirri}},
  \bibinfo {author} {\bibfnamefont {M.}~\bibnamefont {Affronte}}, \bibinfo
  {author} {\bibfnamefont {S.}~\bibnamefont {Carretta}}, \bibinfo {author}
  {\bibfnamefont {P.}~\bibnamefont {Santini}}, \bibinfo {author} {\bibfnamefont
  {G.}~\bibnamefont {Amoretti}}, \bibinfo {author} {\bibfnamefont
  {S.}~\bibnamefont {Piligkos}}, \bibinfo {author} {\bibfnamefont
  {G.}~\bibnamefont {Timco}}, \ and\ \bibinfo {author} {\bibfnamefont
  {R.~E.~P.}\ \bibnamefont {Winpenny}},\ }\href {\doibase
  10.1103/PhysRevLett.94.207208} {\bibfield  {journal} {\bibinfo  {journal}
  {Phys. Rev. Lett.}\ }\textbf {\bibinfo {volume} {94}},\ \bibinfo {pages}
  {207208} (\bibinfo {year} {2005})}\BibitemShut {NoStop}%
\bibitem [{\citenamefont {Bogani}\ and\ \citenamefont
  {Wernsdorfer}(2008)}]{Bogani2008}%
  \BibitemOpen
  \bibfield  {author} {\bibinfo {author} {\bibfnamefont {L.}~\bibnamefont
  {Bogani}}\ and\ \bibinfo {author} {\bibfnamefont {W.}~\bibnamefont
  {Wernsdorfer}},\ }\href {\doibase 10.1038/nmat2133} {\bibfield  {journal}
  {\bibinfo  {journal} {Nat. Mater.}\ }\textbf {\bibinfo {volume} {7}},\
  \bibinfo {pages} {179} (\bibinfo {year} {2008})}\BibitemShut {NoStop}%
\bibitem [{\citenamefont {Loth}\ \emph {et~al.}(2012)\citenamefont {Loth},
  \citenamefont {Baumann}, \citenamefont {Lutz}, \citenamefont {Eigler},\ and\
  \citenamefont {Heinrich}}]{Loth2012}%
  \BibitemOpen
  \bibfield  {author} {\bibinfo {author} {\bibfnamefont {S.}~\bibnamefont
  {Loth}}, \bibinfo {author} {\bibfnamefont {S.}~\bibnamefont {Baumann}},
  \bibinfo {author} {\bibfnamefont {C.~P.}\ \bibnamefont {Lutz}}, \bibinfo
  {author} {\bibfnamefont {D.~M.}\ \bibnamefont {Eigler}}, \ and\ \bibinfo
  {author} {\bibfnamefont {A.~J.}\ \bibnamefont {Heinrich}},\ }\href {\doibase
  10.1126/science.1214131} {\bibfield  {journal} {\bibinfo  {journal}
  {Science}\ }\textbf {\bibinfo {volume} {335}},\ \bibinfo {pages} {196}
  (\bibinfo {year} {2012})}\BibitemShut {NoStop}%
\bibitem [{\citenamefont {Heinrich}\ \emph {et~al.}(2013)\citenamefont
  {Heinrich}, \citenamefont {Braun}, \citenamefont {Pascual},\ and\
  \citenamefont {Franke}}]{Heinrich2013}%
  \BibitemOpen
  \bibfield  {author} {\bibinfo {author} {\bibfnamefont {B.~W.}\ \bibnamefont
  {Heinrich}}, \bibinfo {author} {\bibfnamefont {L.}~\bibnamefont {Braun}},
  \bibinfo {author} {\bibfnamefont {J.~I.}\ \bibnamefont {Pascual}}, \ and\
  \bibinfo {author} {\bibfnamefont {K.~J.}\ \bibnamefont {Franke}},\ }\href
  {\doibase 10.1038/NPHYS2794} {\bibfield  {journal} {\bibinfo  {journal} {Nat.
  Phys.}\ }\textbf {\bibinfo {volume} {9}},\ \bibinfo {pages} {765} (\bibinfo
  {year} {2013})}\BibitemShut {NoStop}%
\bibitem [{\citenamefont {Bacon}\ \emph {et~al.}(2001)\citenamefont {Bacon},
  \citenamefont {Brown},\ and\ \citenamefont {Whaley}}]{Bacon2001}%
  \BibitemOpen
  \bibfield  {author} {\bibinfo {author} {\bibfnamefont {D.}~\bibnamefont
  {Bacon}}, \bibinfo {author} {\bibfnamefont {K.~R.}\ \bibnamefont {Brown}}, \
  and\ \bibinfo {author} {\bibfnamefont {K.~B.}\ \bibnamefont {Whaley}},\
  }\href {\doibase 10.1103/PhysRevLett.87.247902} {\bibfield  {journal}
  {\bibinfo  {journal} {Phys. Rev. Lett.}\ }\textbf {\bibinfo {volume} {87}},\
  \bibinfo {pages} {247902} (\bibinfo {year} {2001})}\BibitemShut {NoStop}%
\bibitem [{\citenamefont {Lidar}\ and\ \citenamefont
  {Whaley}(2003)}]{Lidar2003}%
  \BibitemOpen
  \bibfield  {author} {\bibinfo {author} {\bibfnamefont {D.~A.}\ \bibnamefont
  {Lidar}}\ and\ \bibinfo {author} {\bibfnamefont {K.~B.}\ \bibnamefont
  {Whaley}},\ }in\ \href {\doibase 10.1007/3-540-44874-8_5} {\emph {\bibinfo
  {booktitle} {Irreversible Quantum Dynamics}}}\ (\bibinfo  {publisher}
  {Springer},\ \bibinfo {address} {Berlin, Heidelberg},\ \bibinfo {year}
  {2003})\ pp.\ \bibinfo {pages} {83--120}\BibitemShut {NoStop}%
\bibitem [{\citenamefont {Lidar}(2014)}]{Lidar2014}%
  \BibitemOpen
  \bibfield  {author} {\bibinfo {author} {\bibfnamefont {D.~A.}\ \bibnamefont
  {Lidar}},\ }in\ \href {\doibase 10.1002/9781118742631.ch11} {\emph {\bibinfo
  {booktitle} {Quantum information and computation for chemistry: advances in
  chemical physics}}},\ Vol.\ \bibinfo {volume} {154}\ (\bibinfo  {publisher}
  {John Wiley {\&} Sons, Inc.},\ \bibinfo {address} {Hoboken, New Jersey},\
  \bibinfo {year} {2014})\ pp.\ \bibinfo {pages} {295--354}\BibitemShut
  {NoStop}%
\bibitem [{\citenamefont {Shakirov}\ \emph {et~al.}(2016)\citenamefont
  {Shakirov}, \citenamefont {Shchadilova}, \citenamefont {Rubtsov},\ and\
  \citenamefont {Ribeiro}}]{Shakirov2016a}%
  \BibitemOpen
  \bibfield  {author} {\bibinfo {author} {\bibfnamefont {A.~M.}\ \bibnamefont
  {Shakirov}}, \bibinfo {author} {\bibfnamefont {Y.~E.}\ \bibnamefont
  {Shchadilova}}, \bibinfo {author} {\bibfnamefont {A.~N.}\ \bibnamefont
  {Rubtsov}}, \ and\ \bibinfo {author} {\bibfnamefont {P.}~\bibnamefont
  {Ribeiro}},\ }\href {\doibase 10.1103/PhysRevB.94.224425} {\bibfield
  {journal} {\bibinfo  {journal} {Phys. Rev. B}\ }\textbf {\bibinfo {volume}
  {94}},\ \bibinfo {pages} {224425} (\bibinfo {year} {2016})}\BibitemShut
  {NoStop}%
\bibitem [{\citenamefont
  {Fern{\'{a}}ndez-Rossier}(2009)}]{Fernandez-Rossier2009}%
  \BibitemOpen
  \bibfield  {author} {\bibinfo {author} {\bibfnamefont {J.}~\bibnamefont
  {Fern{\'{a}}ndez-Rossier}},\ }\href {\doibase 10.1103/PhysRevLett.102.256802}
  {\bibfield  {journal} {\bibinfo  {journal} {Phys. Rev. Lett.}\ }\textbf
  {\bibinfo {volume} {102}},\ \bibinfo {pages} {256802} (\bibinfo {year}
  {2009})}\BibitemShut {NoStop}%
\bibitem [{\citenamefont {Lorente}\ and\ \citenamefont
  {Gauyacq}(2009)}]{Lorente2009}%
  \BibitemOpen
  \bibfield  {author} {\bibinfo {author} {\bibfnamefont {N.}~\bibnamefont
  {Lorente}}\ and\ \bibinfo {author} {\bibfnamefont {J.-P.}\ \bibnamefont
  {Gauyacq}},\ }\href {\doibase 10.1103/PhysRevLett.103.176601} {\bibfield
  {journal} {\bibinfo  {journal} {Phys. Rev. Lett.}\ }\textbf {\bibinfo
  {volume} {103}},\ \bibinfo {pages} {176601} (\bibinfo {year}
  {2009})}\BibitemShut {NoStop}%
\bibitem [{\citenamefont {Ternes}(2015)}]{Ternes2015}%
  \BibitemOpen
  \bibfield  {author} {\bibinfo {author} {\bibfnamefont {M.}~\bibnamefont
  {Ternes}},\ }\href {\doibase 10.1088/1367-2630/17/6/063016} {\bibfield
  {journal} {\bibinfo  {journal} {New J. Phys.}\ }\textbf {\bibinfo {volume}
  {17}},\ \bibinfo {pages} {063016} (\bibinfo {year} {2015})}\BibitemShut
  {NoStop}%
\bibitem [{\citenamefont {Anderson}(1966)}]{Anderson1966}%
  \BibitemOpen
  \bibfield  {author} {\bibinfo {author} {\bibfnamefont {P.~W.}\ \bibnamefont
  {Anderson}},\ }\href {\doibase 10.1103/PhysRevLett.17.95} {\bibfield
  {journal} {\bibinfo  {journal} {Phys. Rev. Lett.}\ }\textbf {\bibinfo
  {volume} {17}},\ \bibinfo {pages} {95} (\bibinfo {year} {1966})}\BibitemShut
  {NoStop}%
\bibitem [{\citenamefont {Schrieffer}\ and\ \citenamefont
  {Wolff}(1966)}]{Schrieffer1966}%
  \BibitemOpen
  \bibfield  {author} {\bibinfo {author} {\bibfnamefont {J.~R.}\ \bibnamefont
  {Schrieffer}}\ and\ \bibinfo {author} {\bibfnamefont {P.~A.}\ \bibnamefont
  {Wolff}},\ }\href {\doibase 10.1103/PhysRev.149.491} {\bibfield  {journal}
  {\bibinfo  {journal} {Phys. Rev.}\ }\textbf {\bibinfo {volume} {149}},\
  \bibinfo {pages} {491} (\bibinfo {year} {1966})}\BibitemShut {NoStop}%
\bibitem [{\citenamefont {Gatteschi}\ \emph {et~al.}(2006)\citenamefont
  {Gatteschi}, \citenamefont {Sessoli},\ and\ \citenamefont
  {Villain}}]{Gatteschi2006}%
  \BibitemOpen
  \bibfield  {author} {\bibinfo {author} {\bibfnamefont {D.}~\bibnamefont
  {Gatteschi}}, \bibinfo {author} {\bibfnamefont {R.}~\bibnamefont {Sessoli}},
  \ and\ \bibinfo {author} {\bibfnamefont {J.}~\bibnamefont {Villain}},\
  }\href@noop {} {\emph {\bibinfo {title} {{Molecular nanomagnets}}}},\
  Vol.~\bibinfo {volume} {5}\ (\bibinfo  {publisher} {Oxford University
  Press},\ \bibinfo {address} {Oxford},\ \bibinfo {year} {2006})\BibitemShut
  {NoStop}%
\bibitem [{\citenamefont {Otte}\ \emph {et~al.}(2008)\citenamefont {Otte},
  \citenamefont {Ternes}, \citenamefont {von Bergmann}, \citenamefont {Loth},
  \citenamefont {Brune}, \citenamefont {Lutz}, \citenamefont {Hirjibehedin},\
  and\ \citenamefont {Heinrich}}]{Otte2008}%
  \BibitemOpen
  \bibfield  {author} {\bibinfo {author} {\bibfnamefont {A.~F.}\ \bibnamefont
  {Otte}}, \bibinfo {author} {\bibfnamefont {M.}~\bibnamefont {Ternes}},
  \bibinfo {author} {\bibfnamefont {K.}~\bibnamefont {von Bergmann}}, \bibinfo
  {author} {\bibfnamefont {S.}~\bibnamefont {Loth}}, \bibinfo {author}
  {\bibfnamefont {H.}~\bibnamefont {Brune}}, \bibinfo {author} {\bibfnamefont
  {C.~P.}\ \bibnamefont {Lutz}}, \bibinfo {author} {\bibfnamefont {C.~F.}\
  \bibnamefont {Hirjibehedin}}, \ and\ \bibinfo {author} {\bibfnamefont
  {A.~J.}\ \bibnamefont {Heinrich}},\ }\href {\doibase 10.1038/nphys1072}
  {\bibfield  {journal} {\bibinfo  {journal} {Nat. Phys.}\ }\textbf {\bibinfo
  {volume} {4}},\ \bibinfo {pages} {847} (\bibinfo {year} {2008})}\BibitemShut
  {NoStop}%
\bibitem [{\citenamefont {Appelbaum}(1967)}]{Appelbaum1967}%
  \BibitemOpen
  \bibfield  {author} {\bibinfo {author} {\bibfnamefont {J.~A.}\ \bibnamefont
  {Appelbaum}},\ }\href {\doibase 10.1103/PhysRev.154.633} {\bibfield
  {journal} {\bibinfo  {journal} {Phys. Rev.}\ }\textbf {\bibinfo {volume}
  {154}},\ \bibinfo {pages} {633} (\bibinfo {year} {1967})}\BibitemShut
  {NoStop}%
\bibitem [{\citenamefont {Kim}\ and\ \citenamefont {Kim}(2004)}]{Kim2004}%
  \BibitemOpen
  \bibfield  {author} {\bibinfo {author} {\bibfnamefont {G.~H.}\ \bibnamefont
  {Kim}}\ and\ \bibinfo {author} {\bibfnamefont {T.~S.}\ \bibnamefont {Kim}},\
  }\href {\doibase 10.1103/PhysRevLett.92.137203} {\bibfield  {journal}
  {\bibinfo  {journal} {Phys. Rev. Lett.}\ }\textbf {\bibinfo {volume} {92}},\
  \bibinfo {pages} {137203} (\bibinfo {year} {2004})}\BibitemShut {NoStop}%
\bibitem [{\citenamefont {Breuer}\ and\ \citenamefont
  {Petruccione}(2002)}]{Breuer2002}%
  \BibitemOpen
  \bibfield  {author} {\bibinfo {author} {\bibfnamefont {H.-P.}\ \bibnamefont
  {Breuer}}\ and\ \bibinfo {author} {\bibfnamefont {F.}~\bibnamefont
  {Petruccione}},\ }\href@noop {} {\emph {\bibinfo {title} {{The theory of open
  quantum systems}}}}\ (\bibinfo  {publisher} {Oxford University Press},\
  \bibinfo {address} {Oxford},\ \bibinfo {year} {2002})\BibitemShut {NoStop}%
\bibitem [{\citenamefont {{Le Gall}}\ \emph {et~al.}(2009)\citenamefont {{Le
  Gall}}, \citenamefont {Besombes}, \citenamefont {Boukari}, \citenamefont
  {Kolodka}, \citenamefont {Cibert},\ and\ \citenamefont
  {Mariette}}]{LeGall2009}%
  \BibitemOpen
  \bibfield  {author} {\bibinfo {author} {\bibfnamefont {C.}~\bibnamefont {{Le
  Gall}}}, \bibinfo {author} {\bibfnamefont {L.}~\bibnamefont {Besombes}},
  \bibinfo {author} {\bibfnamefont {H.}~\bibnamefont {Boukari}}, \bibinfo
  {author} {\bibfnamefont {R.}~\bibnamefont {Kolodka}}, \bibinfo {author}
  {\bibfnamefont {J.}~\bibnamefont {Cibert}}, \ and\ \bibinfo {author}
  {\bibfnamefont {H.}~\bibnamefont {Mariette}},\ }\href {\doibase
  10.1103/PhysRevLett.102.127402} {\bibfield  {journal} {\bibinfo  {journal}
  {Phys. Rev. Lett.}\ }\textbf {\bibinfo {volume} {102}},\ \bibinfo {pages}
  {127402} (\bibinfo {year} {2009})}\BibitemShut {NoStop}%
\bibitem [{\citenamefont {Wichterich}\ \emph {et~al.}(2007)\citenamefont
  {Wichterich}, \citenamefont {Henrich}, \citenamefont {Breuer}, \citenamefont
  {Gemmer},\ and\ \citenamefont {Michel}}]{Wichterich2007}%
  \BibitemOpen
  \bibfield  {author} {\bibinfo {author} {\bibfnamefont {H.}~\bibnamefont
  {Wichterich}}, \bibinfo {author} {\bibfnamefont {M.~J.}\ \bibnamefont
  {Henrich}}, \bibinfo {author} {\bibfnamefont {H.-p.}\ \bibnamefont {Breuer}},
  \bibinfo {author} {\bibfnamefont {J.}~\bibnamefont {Gemmer}}, \ and\ \bibinfo
  {author} {\bibfnamefont {M.}~\bibnamefont {Michel}},\ }\href {\doibase
  10.1103/PhysRevE.76.031115} {\bibfield  {journal} {\bibinfo  {journal} {Phys.
  Rev. E}\ }\textbf {\bibinfo {volume} {76}},\ \bibinfo {pages} {031115}
  (\bibinfo {year} {2007})}\BibitemShut {NoStop}%
\bibitem [{\citenamefont {Slonczewski}(1996)}]{Slonczewski1996}%
  \BibitemOpen
  \bibfield  {author} {\bibinfo {author} {\bibfnamefont {J.}~\bibnamefont
  {Slonczewski}},\ }\href {\doibase 10.1016/0304-8853(96)00062-5} {\bibfield
  {journal} {\bibinfo  {journal} {J. Magn. Magn. Mater.}\ }\textbf {\bibinfo
  {volume} {159}},\ \bibinfo {pages} {L1} (\bibinfo {year} {1996})}\BibitemShut
  {NoStop}%
\bibitem [{\citenamefont {Delgado}\ \emph {et~al.}(2010)\citenamefont
  {Delgado}, \citenamefont {Palacios},\ and\ \citenamefont
  {Fern{\'{a}}ndez-Rossier}}]{Delgado2010}%
  \BibitemOpen
  \bibfield  {author} {\bibinfo {author} {\bibfnamefont {F.}~\bibnamefont
  {Delgado}}, \bibinfo {author} {\bibfnamefont {J.~J.}\ \bibnamefont
  {Palacios}}, \ and\ \bibinfo {author} {\bibfnamefont {J.}~\bibnamefont
  {Fern{\'{a}}ndez-Rossier}},\ }\href {\doibase 10.1103/PhysRevLett.104.026601}
  {\bibfield  {journal} {\bibinfo  {journal} {Phys. Rev. Lett.}\ }\textbf
  {\bibinfo {volume} {104}},\ \bibinfo {pages} {026601} (\bibinfo {year}
  {2010})}\BibitemShut {NoStop}%
\bibitem [{\citenamefont {Uchiyama}\ \emph {et~al.}(2009)\citenamefont
  {Uchiyama}, \citenamefont {Aihara}, \citenamefont {Saeki},\ and\
  \citenamefont {Miyashita}}]{Uchiyama2009}%
  \BibitemOpen
  \bibfield  {author} {\bibinfo {author} {\bibfnamefont {C.}~\bibnamefont
  {Uchiyama}}, \bibinfo {author} {\bibfnamefont {M.}~\bibnamefont {Aihara}},
  \bibinfo {author} {\bibfnamefont {M.}~\bibnamefont {Saeki}}, \ and\ \bibinfo
  {author} {\bibfnamefont {S.}~\bibnamefont {Miyashita}},\ }\href {\doibase
  10.1103/PhysRevE.80.021128} {\bibfield  {journal} {\bibinfo  {journal} {Phys.
  Rev. E}\ }\textbf {\bibinfo {volume} {80}},\ \bibinfo {pages} {021128}
  (\bibinfo {year} {2009})}\BibitemShut {NoStop}%
\bibitem [{\citenamefont {Oberg}\ \emph {et~al.}(2014)\citenamefont {Oberg},
  \citenamefont {Calvo}, \citenamefont {Delgado}, \citenamefont {Moro-Lagares},
  \citenamefont {Serrate}, \citenamefont {Jacob}, \citenamefont
  {Fern{\'{a}}ndez-Rossier},\ and\ \citenamefont {Hirjibehedin}}]{Oberg2014}%
  \BibitemOpen
  \bibfield  {author} {\bibinfo {author} {\bibfnamefont {J.~C.}\ \bibnamefont
  {Oberg}}, \bibinfo {author} {\bibfnamefont {M.~R.}\ \bibnamefont {Calvo}},
  \bibinfo {author} {\bibfnamefont {F.}~\bibnamefont {Delgado}}, \bibinfo
  {author} {\bibfnamefont {M.}~\bibnamefont {Moro-Lagares}}, \bibinfo {author}
  {\bibfnamefont {D.}~\bibnamefont {Serrate}}, \bibinfo {author} {\bibfnamefont
  {D.}~\bibnamefont {Jacob}}, \bibinfo {author} {\bibfnamefont
  {J.}~\bibnamefont {Fern{\'{a}}ndez-Rossier}}, \ and\ \bibinfo {author}
  {\bibfnamefont {C.~F.}\ \bibnamefont {Hirjibehedin}},\ }\href {\doibase
  10.1038/nnano.2013.264} {\bibfield  {journal} {\bibinfo  {journal} {Nat.
  Nanotechnol.}\ }\textbf {\bibinfo {volume} {9}},\ \bibinfo {pages} {64}
  (\bibinfo {year} {2014})}\BibitemShut {NoStop}%
\bibitem [{\citenamefont {Bloch}(1946)}]{Bloch1946}%
  \BibitemOpen
  \bibfield  {author} {\bibinfo {author} {\bibfnamefont {F.}~\bibnamefont
  {Bloch}},\ }\href {\doibase 10.1103/PhysRev.70.460} {\bibfield  {journal}
  {\bibinfo  {journal} {Phys. Rev.}\ }\textbf {\bibinfo {volume} {70}},\
  \bibinfo {pages} {460} (\bibinfo {year} {1946})}\BibitemShut {NoStop}%
\bibitem [{\citenamefont {Cohen-Tannoudji}\ \emph {et~al.}(1992)\citenamefont
  {Cohen-Tannoudji}, \citenamefont {Dupont-Roc}, \citenamefont {Grynberg},\
  and\ \citenamefont {Thickstun}}]{Cohen-Tannoudji1992}%
  \BibitemOpen
  \bibfield  {author} {\bibinfo {author} {\bibfnamefont {C.}~\bibnamefont
  {Cohen-Tannoudji}}, \bibinfo {author} {\bibfnamefont {J.}~\bibnamefont
  {Dupont-Roc}}, \bibinfo {author} {\bibfnamefont {G.}~\bibnamefont
  {Grynberg}}, \ and\ \bibinfo {author} {\bibfnamefont {P.}~\bibnamefont
  {Thickstun}},\ }\href@noop {} {\emph {\bibinfo {title} {{Atom-photon
  interactions: basic processes and applications}}}}\ (\bibinfo  {publisher}
  {Wiley},\ \bibinfo {address} {New York},\ \bibinfo {year} {1992})\BibitemShut
  {NoStop}%
\bibitem [{\citenamefont {Kimura}(2002)}]{Kimura2002}%
  \BibitemOpen
  \bibfield  {author} {\bibinfo {author} {\bibfnamefont {G.}~\bibnamefont
  {Kimura}},\ }\href {\doibase 10.1103/PhysRevA.66.062113} {\bibfield
  {journal} {\bibinfo  {journal} {Phys. Rev. A}\ }\textbf {\bibinfo {volume}
  {66}},\ \bibinfo {pages} {062113} (\bibinfo {year} {2002})}\BibitemShut
  {NoStop}%
\bibitem [{\citenamefont {Sudarshan}(2003)}]{Sudarshan2003}%
  \BibitemOpen
  \bibfield  {author} {\bibinfo {author} {\bibfnamefont {E.~C.~G.}\
  \bibnamefont {Sudarshan}},\ }\href {\doibase 10.1016/S0960-0779(02)00297-7}
  {\bibfield  {journal} {\bibinfo  {journal} {Chaos, Solitons {\&} Fractals}\
  }\textbf {\bibinfo {volume} {16}},\ \bibinfo {pages} {369} (\bibinfo {year}
  {2003})}\BibitemShut {NoStop}%
\bibitem [{\citenamefont {Khajetoorians}\ \emph {et~al.}(2013)\citenamefont
  {Khajetoorians}, \citenamefont {Baxevanis}, \citenamefont {H{\"{u}}bner},
  \citenamefont {Schlenk}, \citenamefont {Krause}, \citenamefont {Wehling},
  \citenamefont {Lounis}, \citenamefont {Lichtenstein}, \citenamefont
  {Pfannkuche}, \citenamefont {Wiebe},\ and\ \citenamefont
  {Wiesendanger}}]{Khajetoorians2013}%
  \BibitemOpen
  \bibfield  {author} {\bibinfo {author} {\bibfnamefont {A.~A.}\ \bibnamefont
  {Khajetoorians}}, \bibinfo {author} {\bibfnamefont {B.}~\bibnamefont
  {Baxevanis}}, \bibinfo {author} {\bibfnamefont {C.}~\bibnamefont
  {H{\"{u}}bner}}, \bibinfo {author} {\bibfnamefont {T.}~\bibnamefont
  {Schlenk}}, \bibinfo {author} {\bibfnamefont {S.}~\bibnamefont {Krause}},
  \bibinfo {author} {\bibfnamefont {T.~O.}\ \bibnamefont {Wehling}}, \bibinfo
  {author} {\bibfnamefont {S.}~\bibnamefont {Lounis}}, \bibinfo {author}
  {\bibfnamefont {A.}~\bibnamefont {Lichtenstein}}, \bibinfo {author}
  {\bibfnamefont {D.}~\bibnamefont {Pfannkuche}}, \bibinfo {author}
  {\bibfnamefont {J.}~\bibnamefont {Wiebe}}, \ and\ \bibinfo {author}
  {\bibfnamefont {R.}~\bibnamefont {Wiesendanger}},\ }\href {\doibase
  10.1126/science.1228519} {\bibfield  {journal} {\bibinfo  {journal}
  {Science}\ }\textbf {\bibinfo {volume} {339}},\ \bibinfo {pages} {55}
  (\bibinfo {year} {2013})}\BibitemShut {NoStop}%
\bibitem [{\citenamefont {Miyamachi}\ \emph {et~al.}(2013)\citenamefont
  {Miyamachi}, \citenamefont {Schuh}, \citenamefont {M{\"{a}}rkl},\ and\
  \citenamefont {Bresch}}]{Miyamachi2013}%
  \BibitemOpen
  \bibfield  {author} {\bibinfo {author} {\bibfnamefont {T.}~\bibnamefont
  {Miyamachi}}, \bibinfo {author} {\bibfnamefont {T.}~\bibnamefont {Schuh}},
  \bibinfo {author} {\bibfnamefont {T.}~\bibnamefont {M{\"{a}}rkl}}, \ and\
  \bibinfo {author} {\bibfnamefont {C.}~\bibnamefont {Bresch}},\ }\href
  {\doibase 10.1038/nature12759} {\bibfield  {journal} {\bibinfo  {journal}
  {Nature (London)}\ }\textbf {\bibinfo {volume} {503}},\ \bibinfo {pages}
  {242} (\bibinfo {year} {2013})}\BibitemShut {NoStop}%
\bibitem [{\citenamefont {Kitaev}(2001)}]{Kitaev2001}%
  \BibitemOpen
  \bibfield  {author} {\bibinfo {author} {\bibfnamefont {A.~Y.}\ \bibnamefont
  {Kitaev}},\ }\href {\doibase 10.1070/1063-7869/44/10S/S29} {\bibfield
  {journal} {\bibinfo  {journal} {Physics-Uspekhi}\ }\textbf {\bibinfo {volume}
  {44}},\ \bibinfo {pages} {131} (\bibinfo {year} {2001})}\BibitemShut
  {NoStop}%
\bibitem [{\citenamefont {Toskovic}\ \emph {et~al.}(2016)\citenamefont
  {Toskovic}, \citenamefont {van~den Berg}, \citenamefont {Spinelli},
  \citenamefont {Eliens}, \citenamefont {van~den Toorn}, \citenamefont
  {Bryant}, \citenamefont {Caux},\ and\ \citenamefont {Otte}}]{Toskovic2016}%
  \BibitemOpen
  \bibfield  {author} {\bibinfo {author} {\bibfnamefont {R.}~\bibnamefont
  {Toskovic}}, \bibinfo {author} {\bibfnamefont {R.}~\bibnamefont {van~den
  Berg}}, \bibinfo {author} {\bibfnamefont {A.}~\bibnamefont {Spinelli}},
  \bibinfo {author} {\bibfnamefont {I.~S.}\ \bibnamefont {Eliens}}, \bibinfo
  {author} {\bibfnamefont {B.}~\bibnamefont {van~den Toorn}}, \bibinfo {author}
  {\bibfnamefont {B.}~\bibnamefont {Bryant}}, \bibinfo {author} {\bibfnamefont
  {J.-S.}\ \bibnamefont {Caux}}, \ and\ \bibinfo {author} {\bibfnamefont
  {A.~F.}\ \bibnamefont {Otte}},\ }\href {\doibase 10.1038/nphys3722}
  {\bibfield  {journal} {\bibinfo  {journal} {Nat. Phys.}\ }\textbf {\bibinfo
  {volume} {12}},\ \bibinfo {pages} {656} (\bibinfo {year} {2016})}\BibitemShut
  {NoStop}%
\bibitem [{\citenamefont {Dennis}\ \emph {et~al.}(2002)\citenamefont {Dennis},
  \citenamefont {Kitaev}, \citenamefont {Landahl},\ and\ \citenamefont
  {Preskill}}]{Dennis2002}%
  \BibitemOpen
  \bibfield  {author} {\bibinfo {author} {\bibfnamefont {E.}~\bibnamefont
  {Dennis}}, \bibinfo {author} {\bibfnamefont {A.}~\bibnamefont {Kitaev}},
  \bibinfo {author} {\bibfnamefont {A.}~\bibnamefont {Landahl}}, \ and\
  \bibinfo {author} {\bibfnamefont {J.}~\bibnamefont {Preskill}},\ }\href
  {\doibase 10.1063/1.1499754} {\bibfield  {journal} {\bibinfo  {journal} {J.
  Math. Phys.}\ }\textbf {\bibinfo {volume} {43}},\ \bibinfo {pages} {4452}
  (\bibinfo {year} {2002})}\BibitemShut {NoStop}%
\bibitem [{\citenamefont {Delgado}\ \emph {et~al.}(2013)\citenamefont
  {Delgado}, \citenamefont {Batista},\ and\ \citenamefont
  {Fern{\'{a}}ndez-Rossier}}]{Delgado2013}%
  \BibitemOpen
  \bibfield  {author} {\bibinfo {author} {\bibfnamefont {F.}~\bibnamefont
  {Delgado}}, \bibinfo {author} {\bibfnamefont {C.~D.}\ \bibnamefont
  {Batista}}, \ and\ \bibinfo {author} {\bibfnamefont {J.}~\bibnamefont
  {Fern{\'{a}}ndez-Rossier}},\ }\href {\doibase 10.1103/PhysRevLett.111.167201}
  {\bibfield  {journal} {\bibinfo  {journal} {Phys. Rev. Lett.}\ }\textbf
  {\bibinfo {volume} {111}},\ \bibinfo {pages} {167201} (\bibinfo {year}
  {2013})}\BibitemShut {NoStop}%
\bibitem [{\citenamefont {Nadj-Perge}\ \emph {et~al.}(2014)\citenamefont
  {Nadj-Perge}, \citenamefont {Drozdov}, \citenamefont {Li}, \citenamefont
  {Chen}, \citenamefont {Jeon}, \citenamefont {Seo}, \citenamefont {MacDonald},
  \citenamefont {Bernevig},\ and\ \citenamefont {Yazdani}}]{Nadj-Perge2014}%
  \BibitemOpen
  \bibfield  {author} {\bibinfo {author} {\bibfnamefont {S.}~\bibnamefont
  {Nadj-Perge}}, \bibinfo {author} {\bibfnamefont {I.~K.}\ \bibnamefont
  {Drozdov}}, \bibinfo {author} {\bibfnamefont {J.}~\bibnamefont {Li}},
  \bibinfo {author} {\bibfnamefont {H.}~\bibnamefont {Chen}}, \bibinfo {author}
  {\bibfnamefont {S.}~\bibnamefont {Jeon}}, \bibinfo {author} {\bibfnamefont
  {J.}~\bibnamefont {Seo}}, \bibinfo {author} {\bibfnamefont {A.~H.}\
  \bibnamefont {MacDonald}}, \bibinfo {author} {\bibfnamefont {B.~A.}\
  \bibnamefont {Bernevig}}, \ and\ \bibinfo {author} {\bibfnamefont
  {A.}~\bibnamefont {Yazdani}},\ }\href {\doibase 10.1126/science.1259327}
  {\bibfield  {journal} {\bibinfo  {journal} {Science}\ }\textbf {\bibinfo
  {volume} {346}},\ \bibinfo {pages} {602} (\bibinfo {year}
  {2014})}\BibitemShut {NoStop}%
\bibitem [{\citenamefont {Romming}\ \emph {et~al.}(2013)\citenamefont
  {Romming}, \citenamefont {Hanneken}, \citenamefont {Menzel}, \citenamefont
  {Bickel}, \citenamefont {Wolter}, \citenamefont {von Bergmann}, \citenamefont
  {Kubetzka},\ and\ \citenamefont {Wiesendanger}}]{Romming2013}%
  \BibitemOpen
  \bibfield  {author} {\bibinfo {author} {\bibfnamefont {N.}~\bibnamefont
  {Romming}}, \bibinfo {author} {\bibfnamefont {C.}~\bibnamefont {Hanneken}},
  \bibinfo {author} {\bibfnamefont {M.}~\bibnamefont {Menzel}}, \bibinfo
  {author} {\bibfnamefont {J.~E.}\ \bibnamefont {Bickel}}, \bibinfo {author}
  {\bibfnamefont {B.}~\bibnamefont {Wolter}}, \bibinfo {author} {\bibfnamefont
  {K.}~\bibnamefont {von Bergmann}}, \bibinfo {author} {\bibfnamefont
  {A.}~\bibnamefont {Kubetzka}}, \ and\ \bibinfo {author} {\bibfnamefont
  {R.}~\bibnamefont {Wiesendanger}},\ }\href {\doibase 10.1126/science.1240573}
  {\bibfield  {journal} {\bibinfo  {journal} {Science}\ }\textbf {\bibinfo
  {volume} {341}},\ \bibinfo {pages} {636} (\bibinfo {year}
  {2013})}\BibitemShut {NoStop}%
\bibitem [{\citenamefont {Nagaosa}\ and\ \citenamefont
  {Tokura}(2013)}]{Nagaosa2013}%
  \BibitemOpen
  \bibfield  {author} {\bibinfo {author} {\bibfnamefont {N.}~\bibnamefont
  {Nagaosa}}\ and\ \bibinfo {author} {\bibfnamefont {Y.}~\bibnamefont
  {Tokura}},\ }\href {\doibase 10.1038/nnano.2013.243} {\bibfield  {journal}
  {\bibinfo  {journal} {Nat. Nanotechnol.}\ }\textbf {\bibinfo {volume} {8}},\
  \bibinfo {pages} {899} (\bibinfo {year} {2013})}\BibitemShut {NoStop}%
\end{thebibliography}%

\end{document}